\documentclass[11pt,a4paper]{amsart}
\usepackage{amsmath,amssymb,a4wide,amsthm,ifthen,hyperref,xcolor,enumerate,enumitem}
\usepackage[utf8]{inputenc}

\usepackage{graphicx}
\usepackage{url}
\usepackage{ulem}
\usepackage{epstopdf}
\usepackage[section]{placeins}
\usepackage{natbib}
\usepackage{flafter} 
\usepackage{multirow}
\newcommand{\bY}{\boldsymbol{Y}}
\newcommand{\bX}{\boldsymbol{X}}
\newcommand{\bE}{\boldsymbol{E}}
\newcommand{\bB}{\boldsymbol{B}}
\begin{document}

\title[Estimation of large block structured covariance matrices]{Estimation of large block structured covariance matrices: 
   Application to ``multi-omic'' approaches to study seed quality}

\date{\today}

\author{M. Perrot-Dockès}
\address{UMR MIA-Paris, AgroParisTech, INRA, Universit\'e 
Paris-Saclay, 75005, Paris, France}
\email{marie.perrot-dockes@agroparistech.fr}
\author{C. L\'evy-Leduc}
\address{UMR MIA-Paris, AgroParisTech, INRA, Universit\'e 
Paris-Saclay, 75005, Paris, France}
\email{celine.levy-leduc@agroparistech.fr}
\author{L. Rajjou}
\address{Institut Jean-Pierre Bourgin, INRA, AgroParisTech, Université Paris-Saclay, 78026, Versailles, France}
\email{loic.rajjou@agroparistech.fr}

\maketitle

\begin{abstract}
Motivated by an application in high-throughput genomics and metabolomics, we propose a novel, efficient and fully data-driven approach for estimating large block structured sparse covariance matrices in the case where
the number of variables is much larger than the number of samples without limiting ourselves to block diagonal matrices. 
Our approach consists in approximating such a covariance matrix by the sum of a low-rank sparse matrix and a diagonal matrix. 
Our methodology also can deal with matrices for which the block structure appears only if the columns and rows are permuted according to an unknown permutation.
Our technique is implemented in the R package \texttt{BlockCov} which is available from the Comprehensive R Archive Network (CRAN) and from GitHub. 
In order to illustrate the statistical and numerical performance of our package some numerical experiments are provided as
well as a thorough comparison with alternative methods. Finally, our
approach is applied to the use of ``multi-omic'' approaches for studying seed quality. 
\end{abstract}

\section{Introduction}

Plant functional genomics refers to the description of the biological function of a single or a group of genes and both the dynamics and the plasticity of genome expression to shape the phenotype. Combining multi-omics such as transcriptomic, proteomic or metabolomic approaches allows us to address in a new light the dimension and the complexity of the different levels of gene expression control and the delicacy of the metabolic regulation of plants under fluctuation environments. Thus, our era marks a real conceptual shift in plant biology where the individual is no longer considered as a simple sum of components but rather as a system with a set of interacting components to maximize its growth, its reproduction and its adaptation. Plant systems biology is therefore defined by multidisciplinary and multi-scale approaches based on the acquisition of a wide range of data as exhaustive as possible. 

In this context, 
it is crucial to propose new methodologies for integrating heterogeneous data explaining the co-regulations/co-accumulations of products of gene expression (mRNA, proteins) and metabolites.
In order to better understand these phenomena, our goal will thus be to propose a new approach for estimating block structured covariance matrix in a high-dimensional framework
where the  dimension of the covariance matrix is much larger than
the sample size.
 In this setting, it is well known that the commonly used sample covariance matrix performs poorly. 
In recent years, researchers have proposed various regularization techniques to consistently
estimate large covariance matrices or the inverse of such matrices, namely precision matrices. To estimate such matrices, one of
the key assumptions made in the literature is that the matrix of interest is sparse, namely
many entries are equal to zero. A number of regularization approaches including banding, tapering, thresholding and $\ell_1$ minimization, have been developed
to estimate large covariance matrices or their inverse such as, for instance, \cite{LEDOIT2004}, \cite{bickel2008}, \cite{banerjee2008model}, \cite{bien_tibshirani_2011} and
\cite{rothman_2012} among many others. For further references, we refer the reader to
\cite{cai2012} and to the review of \cite{fan_overview_2016}.

In this paper, we shall consider the following framework. Let  $\boldsymbol{E}_1, \boldsymbol{E}_2,\cdots,\boldsymbol{E}_n$, $n$ zero-mean i.i.d. $q$-dimensional 
random vectors having a covariance matrix 
$\boldsymbol{\Sigma}$ such that the number $q$ of its rows and columns is much larger than $n$. 
The goal of the paper is to propose a new estimator of $\boldsymbol{\Sigma}$ and of the square root of its inverse, $\boldsymbol{\Sigma}^{-1/2}$, in the particular case 
where $\boldsymbol{\Sigma}$ is assumed to have a block structure without limiting ourselves to diagonal blocks.
An accurate estimator of $\boldsymbol{\Sigma}$ can indeed be very useful to better understand the links between the columns of the observation matrix and may highlight some biological processes. 
Moreover, an estimator of $\boldsymbol{\Sigma}^{-1/2}$ can be very useful in the general linear model in order to remove the dependence that may exist between the columns 
of the observation matrix. For further details on this point, we refer the reader to \cite{PERROTDOCKES201878}, \cite{perrot2018variable} and to the R package \texttt{MultiVarSel} 
in which such an approach is proposed and implemented for performing variable selection in the multivariate linear model
in the presence of dependence between the columns of the observation matrix.

\begin{figure}[!h]
\begin{center}
\includegraphics[width=0.85\textwidth]{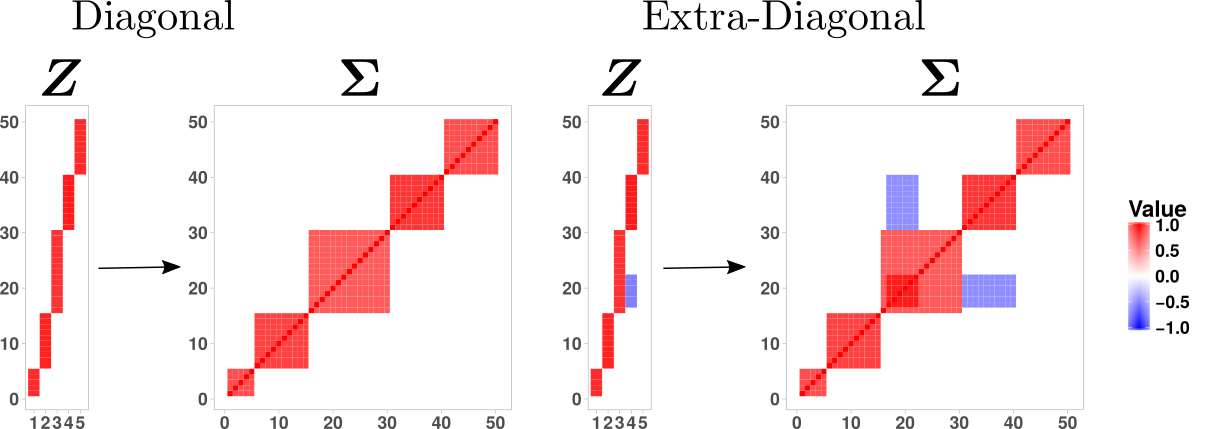}
\caption{Examples of matrices $\boldsymbol{\Sigma}$ generated from different matrices $\boldsymbol{Z}$ leading to a block diagonal or to a more general block structure (extra-diagonal blocks).
\label{fig:Z2sigma}}
\end{center}
\end{figure}

More precisely, in this paper, we shall assume that
\begin{equation}\label{eq:Sigma}
\boldsymbol{\Sigma}=\boldsymbol{Z}\boldsymbol{Z}'+\boldsymbol{D},
\end{equation}
where $\boldsymbol{Z}$ is a $q \times k$ sparse matrix with $k\ll q$, $\boldsymbol{Z}'$ denotes the transpose of the matrix $\boldsymbol{Z}$ and $\boldsymbol{D}$
is a diagonal matrix such that the diagonal terms of $\boldsymbol{\Sigma}$ are equal to one. Two examples of such matrices $\boldsymbol{Z}$ and $\boldsymbol{\Sigma}$ are given
in Figure \ref{fig:Z2sigma} in the case where $k=5$ and $q=50$ and in the case where the columns of $\boldsymbol{\Sigma}$ do not need to be permuted in order
to see the block structure. Based on (\ref{eq:Sigma}), our model could seem to be close to factor models described in \cite{Johnson_1988} and \cite{fan_overview_2016}. 
However, in \cite{Johnson_1988}, the high-dimensional aspects are not considered and in \cite{fan_overview_2016} the sparsity constraint is not studied.
\cite{blum} proposed a methodology which is based on the factor model but with a sparsity constraint on the coefficients of $\boldsymbol{Z}$ which leads to a sparse covariance matrix.
Note also that the block diagonal assumption has already been recently considered by \cite{devijver_2018} for estimating the inverse of large covariance matrices in high-dimensional Gaussian
Graphical Models (GGM).

\begin{figure}[!h]
\begin{center}
\includegraphics[width=0.85\textwidth]{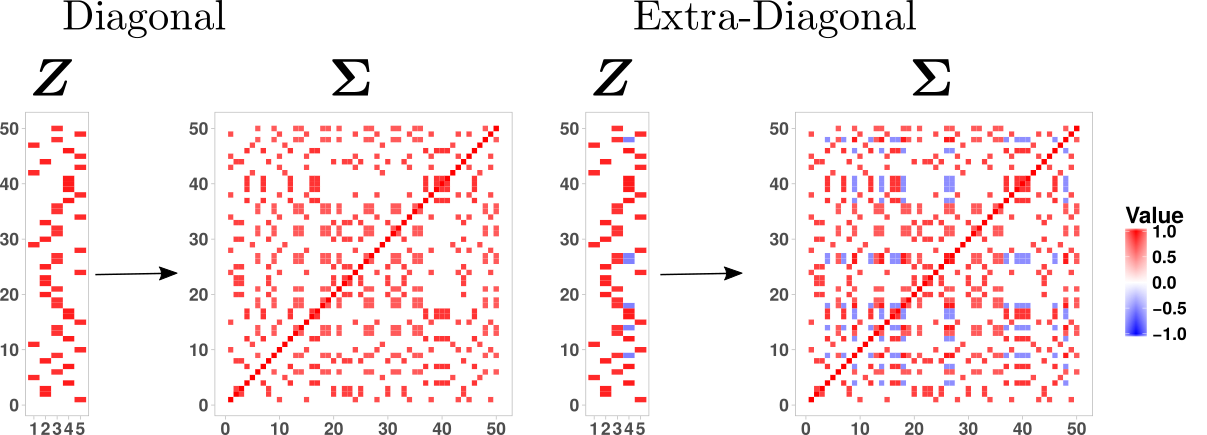}
\caption{Examples of matrices $\boldsymbol{\Sigma}$ of Figure \ref{fig:Z2sigma} in which the columns and rows are randomly permuted.\label{fig:sigma_perm}}
\end{center}
\end{figure}

We also propose a methodology to estimate $\boldsymbol{\Sigma}$ in the case where the block structure is latent; that is, permuting the columns and rows of $\boldsymbol{\Sigma}$ 
renders visible its block structure. An example of such a matrix $\boldsymbol{\Sigma}$ is given
in Figure \ref{fig:sigma_perm} in the case where $k=5$ and $q=50$.

Our approach is fully data-driven and consists in providing a low rank matrix approximation of the $\boldsymbol{Z}\boldsymbol{Z}'$ part of
$\boldsymbol{\Sigma}$ and then in using a $\ell_1$ regularization to obtain a sparse estimator of $\boldsymbol{\Sigma}$.
When the block structure is latent, a hierarchical clustering step must be applied first. With this estimator of $\boldsymbol{\Sigma}$, we explain how
to obtain an estimator of $\boldsymbol{\Sigma}^{-1/2}$.

Our methodology is described in Section \ref{sec:stat_inf}. Some numerical experiments on synthetic data are provided
in Section \ref{sec:num_exp}. An application to the analysis of ``-omic'' data  to study seed quality is performed in Section \ref{sec:real_data}.

\section{Statistical inference}\label{sec:stat_inf}

The strategy that we propose for estimating $\boldsymbol{\Sigma}$ and $\boldsymbol{\Sigma}^{-1/2}$ can be summarized as follows.

 \begin{itemize}
 \item \textsf{First step: Low rank approximation.} In this step, we propose to approximate the part $\boldsymbol{Z}\boldsymbol{Z}'$ of
$\boldsymbol{\Sigma}$ by a low rank matrix using a Singular Value Decomposition (SVD).
 \item \textsf{Second step: Detecting the position of the non null values.} In this step, we use a Lasso criterion to yield a sparse
estimator $\widetilde{\boldsymbol{\Sigma}}$ of $\boldsymbol{\Sigma}$.
\item \textsf{Third step: Positive definiteness.} We apply the methodology of \cite{Higham_2002} to $\widetilde{\boldsymbol{\Sigma}}$ to ensure that the final estimator $\widehat{\boldsymbol{\Sigma}}$ of 
$\boldsymbol{\Sigma}$ is positive definite.
\item \textsf{Fourth step: Estimation of $\boldsymbol{\Sigma}^{-1/2}$.} In this step, $\boldsymbol{\Sigma}^{-1/2}$ is estimated from the spectral decomposition
of $\widehat{\boldsymbol{\Sigma}}$ obtained in the previous step.
\end{itemize}

\subsection{Low rank approximation}\label{ssec:stat_inf_lr}

By definition of $\boldsymbol{Z}$ in (\ref{eq:Sigma}), $\boldsymbol{Z}\boldsymbol{Z}'$ is a $q\times q$ low rank matrix having its rank smaller or equal to $k\ll q$.
In the first step, our goal is thus to propose a low rank approximation of an estimator of $\boldsymbol{Z}\boldsymbol{Z}'$.

Let $\boldsymbol{S}$ be the sample $q\times q$ covariance matrix defined by
$$
\boldsymbol{S}=\frac{1}{n-1}\sum_{i=1}^n  \left(\boldsymbol{E}_{i}-\overline{\boldsymbol{E}}\right)
\left(\boldsymbol{E}_{i}-\overline{\boldsymbol{E}}\right)',\quad\textrm{with } \overline{\boldsymbol{E}}=\frac{1}{n}\sum_{i=1}^n \boldsymbol{E}_{i},
$$
where $\boldsymbol{E}_{i}=(E_{i,1},\dots,E_{i,q})'$. The corresponding $q\times q$ sample correlation matrix $\boldsymbol{R}=(R_{i,j})$ is defined by:
\begin{equation}\label{eq:mat_corr}
R_{i,j}=\frac{S_{i,j}}{\sigma_{i}\sigma_{j}},\; \forall 1\leq i,j\leq q,
\end{equation}
where
$$
\sigma_{i}^2=\frac{1}{n-1}\sum_{\ell=1}^n (E_{\ell,i}-\overline{E}_i)^2,\quad\textrm{with } \overline{E}_i=\frac{1}{n}\sum_{\ell=1}^n E_{\ell,i},\; \forall 1\leq i\leq q.
$$
Let us also consider the $(q-1)\times (q-1)$ matrix $\boldsymbol{\Gamma}$ defined by:
\begin{align}\label{eq:def_Gamma}
\Gamma_{i,j}&=R_{i,j+1},\; \forall 1\leq i\leq j\leq q-1,\\\nonumber
\Gamma_{i,j}&=\Gamma_{j,i},\; \forall 1\leq j< i\leq q-1.
\end{align}
If $\boldsymbol{S}$ was the real matrix $\boldsymbol{\Sigma}$, the corresponding matrix $\Gamma$ would have a rank less than or equal to $k$. Since $\boldsymbol{S}$ is an estimator
of $\boldsymbol{\Sigma}$, we shall use a rank $r$ approximation $\boldsymbol{\Gamma}_r$ of $\boldsymbol{\Gamma}$. This will be performed by considering in its singular value 
decomposition only the $r$ largest singular values
and by replacing the other ones by 0. By \cite{eckart_young_1936}, this corresponds to the best rank $r$ approximation of $\boldsymbol{\Gamma}$. 
The choice of $r$ will be discussed in Section \ref{sec:choix_param}.


\subsection{Detecting the position of the non null values}\label{sec:def_sigma_tilde}

Let us first explain the usual framework in which the Lasso
approach is used. We consider a linear model of the following form
\begin{equation}\label{eq:model_vec}
\mathcal{Y}=\mathcal{X}\mathcal{B}+\mathcal{E},
\end{equation}
where $\mathcal{Y}$, $\mathcal{B}$ and $\mathcal{E}$ are vectors and $\mathcal{B}$ is sparse meaning that it has a lot of null components.

In such models a very popular approach initially proposed by \cite{Tib96} is the Least Absolute
Shrinkage eStimatOr (Lasso), which is defined as follows for a positive $\lambda$:
\begin{equation}\label{eq:lasso}
\widehat{\mathcal{B}}(\lambda)=\textrm{Argmin}_\mathcal{B}\left\{\|\mathcal{Y}-\mathcal{X}\mathcal{B}\|_2^2+\lambda\|\mathcal{B}\|_1\right\},
\end{equation}
where, for $u=(u_1,\dots,u_n)$, $\|u\|_2^2=\sum_{i=1}^n u_i^2$ and $\|u\|_1=\sum_{i=1}^n |u_i|$, \textit{i.e.} the $\ell_1$-norm of the vector $u$. Observe that the first term of (\ref{eq:lasso})
is the classical least-squares criterion and that $\lambda\|\mathcal{B}\|_1$ can be seen as a penalty term. The interest of such
a criterion is the sparsity enforcing property of the  $\ell_1$-norm ensuring that the number of non-zero components of the estimator 
$\widehat{\mathcal{B}}$ of $\mathcal{B}$ is small for large enough values of $\lambda$.
Let 
\begin{equation}\label{eq:def_Y}
\mathcal{Y} = vec_H(\boldsymbol{\Gamma}_r),
\end{equation}
where $vec_H$ defined in Section 16.4 of \cite{harville2001matrix} is such that for a $n\times n$ matrix $A$,
\begin{equation*}
vec_H(A)= \begin{pmatrix}
a_1* \\ a_2*\\ \vdots \\ a_n*
\end{pmatrix},
\end{equation*}
where $a_i*$ is the sub-vector of the column $i$ of $A$ obtained by striking out the $i-1$ first elements.
In order to estimate the sparse matrix $\boldsymbol{Z}\boldsymbol{Z}'$, we need to propose a sparse estimator of $\boldsymbol{\Gamma}_r$. 
To do this we apply the Lasso criterion described in (\ref{eq:lasso}), where $\mathcal{X}$ is the identity matrix.
In the case where $\mathcal{X}$ is an orthogonal matrix it has been shown in \cite{giraud2014introduction} that the solution of \eqref{eq:lasso} is:
\begin{equation*}
\widehat{\mathcal{B}}(\lambda)_j =\left\{
\begin{tabular}{cl}
$\mathcal{X}_j'\mathcal{Y}(1 - \frac{\lambda}{2|\mathcal{X}_j'\mathcal{Y}|} )$, & if  $|\mathcal{X}_j'\mathcal{Y}| > \frac{\lambda}{2}$ \\
$0$, & otherwise,
\end{tabular}
\right.
\end{equation*}
where $\mathcal{X}_j$ denotes the $j$th column of $\mathcal{X}$. Using the fact that $\mathcal{X}$ is the identity matrix we get 
\begin{equation}\label{eq:B_chap}
\widehat{\mathcal{B}}(\lambda)_j =\left\{
\begin{tabular}{cl}
$\mathcal{Y}_j(1 - \frac{\lambda}{2|\mathcal{Y}_j|} )$, & if  $|\mathcal{Y}_j| > \frac{\lambda}{2}$ \\
$0$, & otherwise.
\end{tabular}
\right.
\end{equation}
We then reestimate the non null coefficients using the least-squares criterion and get: 
\begin{equation}\label{eq:B_tilde}
\widetilde{\mathcal{B}}(\lambda)_j =\left\{
\begin{tabular}{cl}
$\mathcal{Y}_j$, & if  $|\mathcal{Y}_j| > \frac{\lambda}{2}$ \\
$0$, & otherwise,
\end{tabular}
\right.
\end{equation}
where $\mathcal{Y}$ is defined in (\ref{eq:def_Y}).

It has to be noticed that $\widehat{\boldsymbol{\Gamma}}_r$ obtained in (\ref{eq:B_chap}) satisfies the following criterion:
$$
\widehat{\boldsymbol{\Gamma}}_r=\textrm{Argmin}_{\boldsymbol{\Theta}}\left\{\|\boldsymbol{\Gamma}_r-\boldsymbol{\Theta}\|_F+\lambda|\boldsymbol{\Theta}|_1\right\},
$$
where $\|\cdot\|_F$ denotes the Frobenius norm defined for a matrix $A$ by $\|A\|_F^2=\textrm{Trace}(A'A)$, 
$|M|_1=\|vec(M)\|_1$ denotes the $\ell_1$-norm of the vector formed by stacking the columns of $M$. It is thus closely related to the 
generalized thresholding estimator defined in \cite{Wen_yang_liu_qiu_2016}
and to the one defined in \cite{rothman_2012} with $\tau=0$ except that in our case $|\boldsymbol{\Theta}^-|_1$ is replaced by $|\boldsymbol{\Theta}|_1$
where  $\boldsymbol{\Theta}^-$ corresponds to the matrix $\boldsymbol{\Theta}$ in which the diagonal terms are replaced by 0. 
The diagonal terms of $\boldsymbol{\Sigma}$ were indeed already removed in $\boldsymbol{\Gamma}_r$.
Hence, we get $\widehat{\boldsymbol{\Gamma}}_r$ by elementwise soft-thresholding that is by putting to zero the value of $\boldsymbol{\Gamma}_r$ that are under a given threshold
and by multiplying the non null values by a coefficient containing this threshold.


Here, we choose to estimate $\boldsymbol{\Gamma}_r$ by $\widetilde{\boldsymbol{\Gamma}}_r(\lambda)$ defined through $\widetilde{\mathcal{B}}(\lambda)$ in (\ref{eq:B_tilde}) which
corresponds to a hard-thresholding and we set the upper triangular part of the estimator $\widetilde{\boldsymbol{\Sigma}}(\lambda)$ of $\boldsymbol{\Sigma}$ 
to be equal to  $\widetilde{\boldsymbol{\Gamma}}_r(\lambda)$. Since the diagonal terms of $\boldsymbol{\Sigma}$ are assumed to be equal to 1, we take the diagonal terms
of $\widetilde{\boldsymbol{\Sigma}}(\lambda)$ equal to 1. The lower triangular part of $\widetilde{\boldsymbol{\Sigma}}(\lambda)$ is then obtained by symmetry. 

The choice of the best parameter $\lambda$ denoted $\lambda_{\textrm{final}}$ in the following will be discussed in Section \ref{sec:lambda}. 

\subsection{Positive definiteness}
To ensure the positive definiteness of our estimator $\widehat{\boldsymbol{\Sigma}}$ of $\boldsymbol{\Sigma}$, we consider
the nearest correlation matrix to $\widetilde{\boldsymbol{\Sigma}}(\lambda_{\textrm{final}})$ which is computed by using the methodology proposed by \cite{Higham_2002} 
and which is implemented in the function \texttt{nearPD} of the R package \texttt{Matrix}, see \cite{Matrix}.  

\subsection{Estimation of $\boldsymbol{\Sigma}^{-1/2}$}
Even if providing an estimator of a large covariance matrix can be very useful in practice, it may also be interesting to efficiently estimate
$\boldsymbol{\Sigma}^{-1/2}$. Such an estimator can indeed be used in the general linear model in order to remove the dependence that may exist between the columns of the observations matrix. For further details on this point, we refer the reader to \cite{PERROTDOCKES201878}, \cite{perrot2018variable} and to the R package \texttt{MultiVarSel} in which such an approach is proposed 
and implemented for performing variable selection in the multivariate linear model
in the presence of dependence between the columns of the observation matrix.

Since $\widehat{\boldsymbol{\Sigma}}$ is a symmetric matrix, it can be rewritten as $\boldsymbol{UDU}'$, where $\boldsymbol{D}$ is a diagonal matrix and 
$\boldsymbol{U}$ is an orthogonal matrix. The matrix $\boldsymbol{\Sigma}^{-1/2}$ can thus be estimated by $\boldsymbol{U}\boldsymbol{D}^{-1/2}\boldsymbol{U}'$ where $\boldsymbol{D}^{-1/2}$ 
is a diagonal matrix
having its diagonal terms equal to the square root of the inverse of the singular values of $\widehat{\boldsymbol{\Sigma}}$.
However, inverting the square root of too small eigenvalues may lead to poor estimators of $\boldsymbol{\Sigma}^{-1/2}$. This is the reason why we propose to estimate $\boldsymbol{\Sigma}^{-1/2}$
by 
\begin{equation}\label{eq:Sigma-1/2_t}
\widehat{\boldsymbol{\Sigma}}^{-1/2}_t=\boldsymbol{U}\boldsymbol{D}_t^{-1/2}\boldsymbol{U}',
\end{equation}
where $\boldsymbol{D}_t^{-1/2}$ is a diagonal matrix such that its diagonal entries are equal to the square root of the inverse of the diagonal entries of $\boldsymbol{D}$ except for those
which are smaller than a given threshold $t$ which are replaced by 0 in $\boldsymbol{D}_t^{-1/2}$. The choice of $t$ will be further discussed in Section \ref{sec:choice_t}.

\subsection{Choice of the parameters}\label{sec:choix_param}
Our methodology for estimating $\boldsymbol{\Sigma}$ depends on two parameters: The number $r$ of singular values kept for defining $\boldsymbol{\Gamma}_r$ 
and the parameter $\lambda$ which controls the sparsity level namely the number of zero values in 
$\widetilde{\mathcal{B}}(\lambda)$ defined in (\ref{eq:B_tilde}).

For choosing $r$, we shall compare two strategies in Section \ref{sec:low_rank}: 
\begin{itemize}
\item The \textsf{Cattell} criterion based on the Cattell's scree plot described in \cite{cattell_1966} and
\item the \textsf{PA} permutation method proposed by \cite{Horn1965} and recently studied from a theoretical point of view by\cite{dobriban_2018}.
\end{itemize} 

To choose the parameter $\lambda$ in (\ref{eq:B_tilde}), we shall compare two strategies in Section \ref{sec:lambda}: 
\begin{itemize}
\item The \textsf{BL} approach proposed in \cite{bickel2008} based on cross-validation and
\item the \textsf{Elbow} method which consists in computing for different values of $\lambda$
 the Frobenius norm $\|\boldsymbol{R}-\widetilde{\boldsymbol{\Sigma}}(\lambda)\|_F$, where $\boldsymbol{R}$ and $\widetilde{\boldsymbol{\Sigma}}(\lambda)$ are  defined
in (\ref{eq:mat_corr}) and at the end of Section \ref{sec:def_sigma_tilde}, respectively. Then, it fits two simple linear regressions and chooses 
the value of $\lambda$ achieving the best fit.
\end{itemize}


\section{Numerical experiments}\label{sec:num_exp}

Our methodology described in the previous section is implemented in the R package \texttt{BlockCov} and is available from the CRAN (Comprehensive R Achive Network)
and from GitHub.

We propose hereafter to investigate the performance of our approach for different types of matrices $\boldsymbol{\Sigma}$ defined in (\ref{eq:Sigma}) and for different values of $n$ and $q$.
The four following cases considered correspond to different types of matrices $\boldsymbol{Z}$, the matrices $\boldsymbol{D}$ being chosen accordingly to ensure that
the matrix $\boldsymbol{\Sigma}$ has its diagonal terms equal to 1.

\begin{itemize}
\item \textbf{Diagonal-Equal} case. In this situation, $\boldsymbol{Z}$ has the structure displayed in the left part of Figure \ref{fig:Z2sigma}, namely it has 5 columns such that
the numbers of the non values in the five columns are equal to $0.1\times q$, $0.2\times q$, $0.3\times q$, $0.2\times q$ and $0.2\times q$, respectively
and the non null values are equal to $\sqrt{0.7}$, $\sqrt{0.75}$, $\sqrt{0.65}$, $\sqrt{0.8}$ and $\sqrt{0.7}$, respectively.
\item \textbf{Diagonal-Unequal} case. In this scenario, $\boldsymbol{Z}$ has the same structure as for the \textbf{Diagonal-Equal} case except that the non null values
in the five columns are not fixed but randomly chosen in $[\sqrt{0.6},\sqrt{0.8}]$ except for the third column for which its values are randomly chosen in $[\sqrt{0.3},\sqrt{0.6}]$.
\item \textbf{Extra-Diagonal-Equal} case. Here, $\boldsymbol{Z}$ has the structure displayed in the right part of Figure \ref{fig:Z2sigma}. The values of the columns of $\boldsymbol{Z}$
are the same as those of the \textbf{Diagonal-Equal} case except for the fourth column which is assumed to contain additional non values equal to -0.5 in the range $[0.35\times q,0.45\times q]$.
\item \textbf{Extra-Diagonal-Unequal} case. $\boldsymbol{Z}$ has the same structure as in the \textbf{Extra-Diagonal-Equal} case except that the values are randomly chosen as in the 
\textbf{Diagonal-Unequal} case except for the fourth column where the additional non values are still equal to -0.5 in the range $[0.35\times q,0.45\times q]$.
\end{itemize}

For $n \in \{10, 30, 50\}$ and $q \in \{100,500\}$, 100 $n\times q$ matrices $\boldsymbol{E}$ were generated such that its rows $\boldsymbol{E}_1, \boldsymbol{E}_2,\cdots,\boldsymbol{E}_n$ 
are i.i.d. $q$-dimensional zero-mean Gaussian vectors having a covariance matrix 
$\boldsymbol{\Sigma}$ chosen according to the four previous cases: \textbf{Diagonal-Equal}, \textbf{Diagonal-Unequal}, \textbf{Extra-Diagonal-Equal}
or \textbf{Extra-Diagonal-Unequal}.

\subsection{Low rank approximation}\label{sec:low_rank}

\begin{figure}
\hspace{-5mm}
\includegraphics[width = \textwidth]{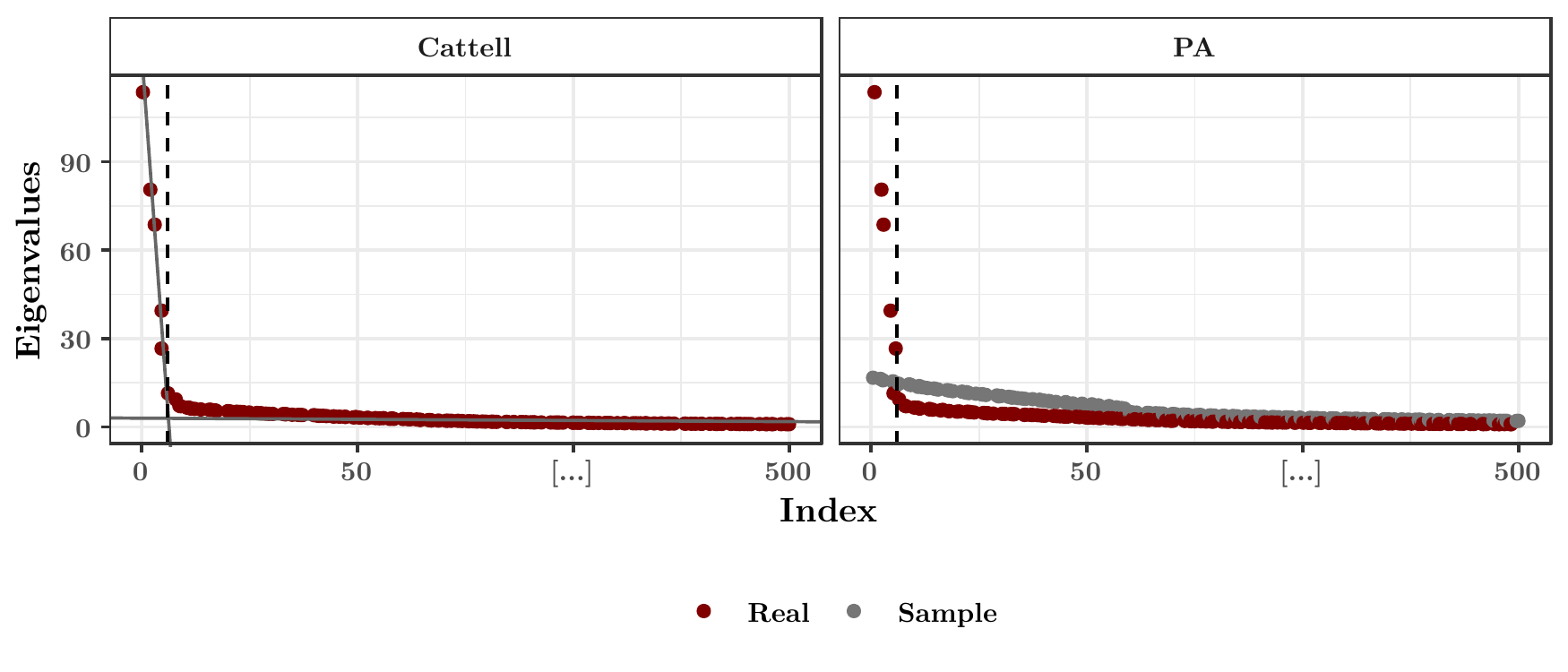}
\caption{Illustration of \textsf{PA} and \textsf{Cattell} criteria for choosing $r$ when $q=500$ and $n=30$ in the Extra-Diagonal-Unequal case.
The value of $r$ found by both methodologies is displayed with a dotted line, the straight lines obtained for the \textsf{Cattell} criterion and the eigenvalues of the permuted matrices in
 the \textsf{PA} methodology are displayed in grey. \label{fig:choixk}}
\end{figure}

The approaches for choosing $r$ described in Section \ref{sec:choix_param} are illustrated in Figure \ref{fig:choixk} in the Extra-Diagonal-Unequal case. We can see from this figure 
that both methodologies find the right value of $r$ which is here equal to 5.

\begin{figure}[!h]
\begin{center}
\includegraphics[width=\textwidth]{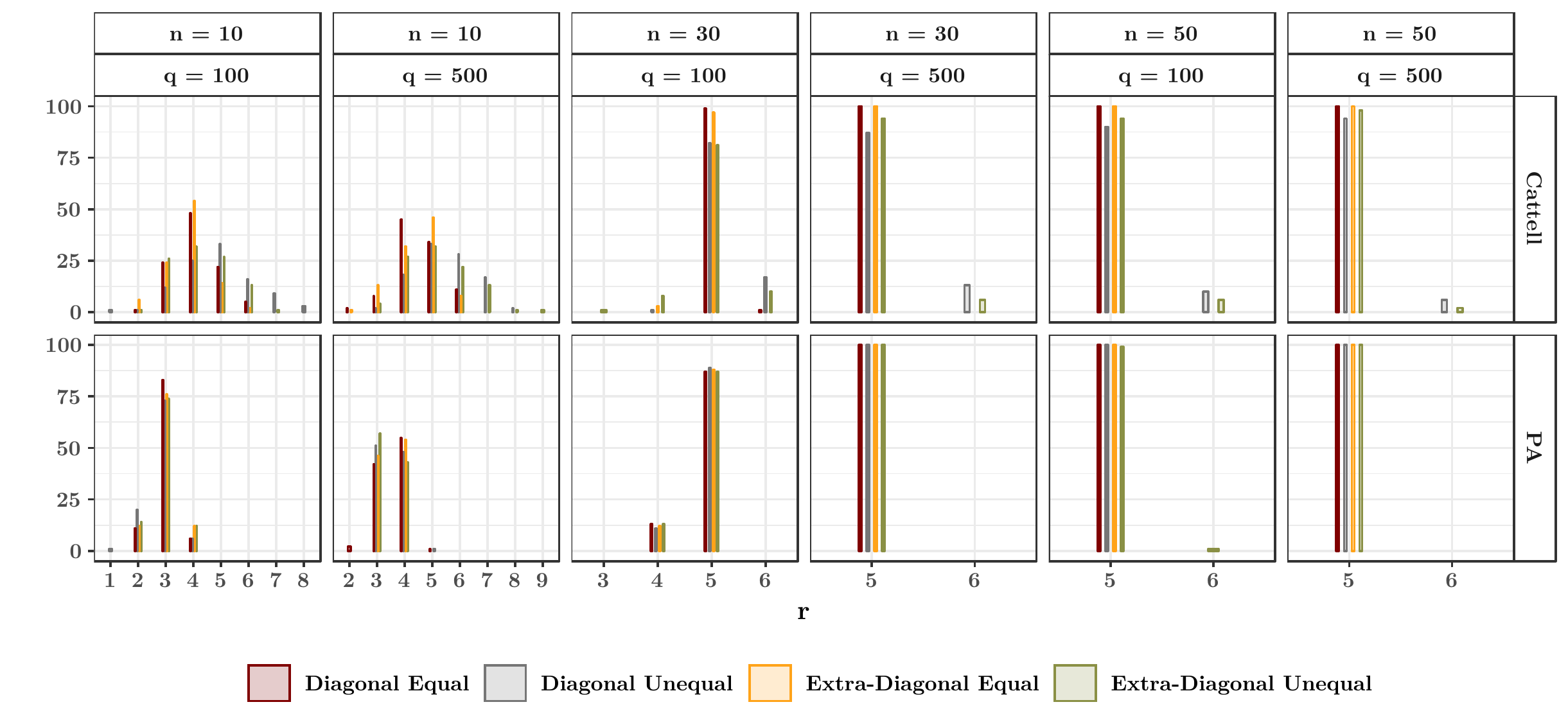}
\caption{Barplots corresponding to the number of times where each value of $r$ is chosen in the low-rank approximation 
from 100 replications for the two methodologies in the different scenarii for the different values of $n$ et $q$.\label{fig:k}}
\end{center}
\end{figure}

\begin{figure}[!h]
\begin{center}
\includegraphics[width=\textwidth]{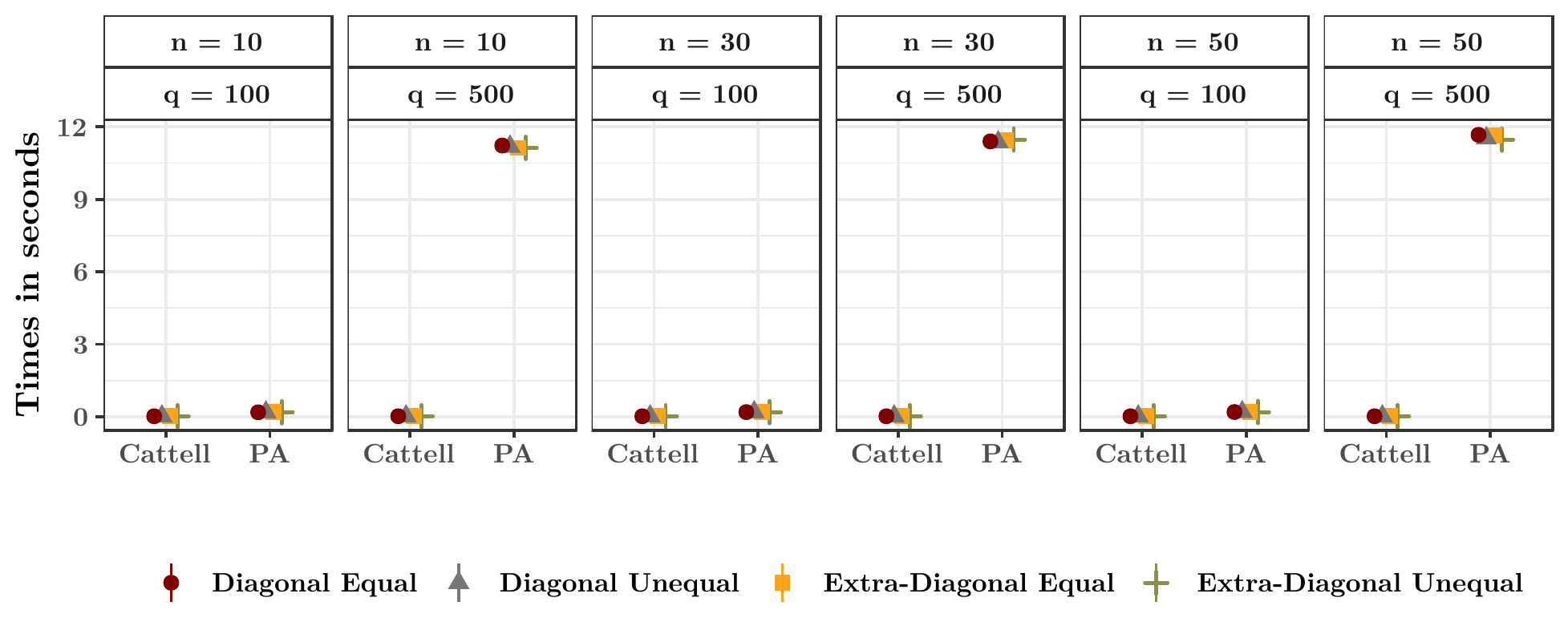}
\caption{Computational times of \textsf{PA} and \textsf{Cattell} criteria. \label{fig:time_k}}
\end{center}
\end{figure}

To go further, we investigate the behavior of our methodologies from 100 replications of the matrix $\boldsymbol{E}$ for the four different types of $\boldsymbol{\Sigma}$.
Figure \ref{fig:k} displays the barplots associated to the estimation of $r$ made in the different replications by the two approaches for the different scenarii.
We can see from this figure that the \textsf{PA} criterion seems to be slightly more stable than the \textsf{Cattell} criterion when $n\geq 30$. However, in the case where $n=10$,  
the \textsf{PA} criterion underestimates the value of $r$. Moreover, in terms of computational time, the performance of \textsf{Cattell} is much better, see Figure \ref{fig:time_k}.

\subsection{Positions of the non null values}\label{sec:lambda}

\begin{figure}[!h]
\begin{center}
\includegraphics[width=\textwidth]{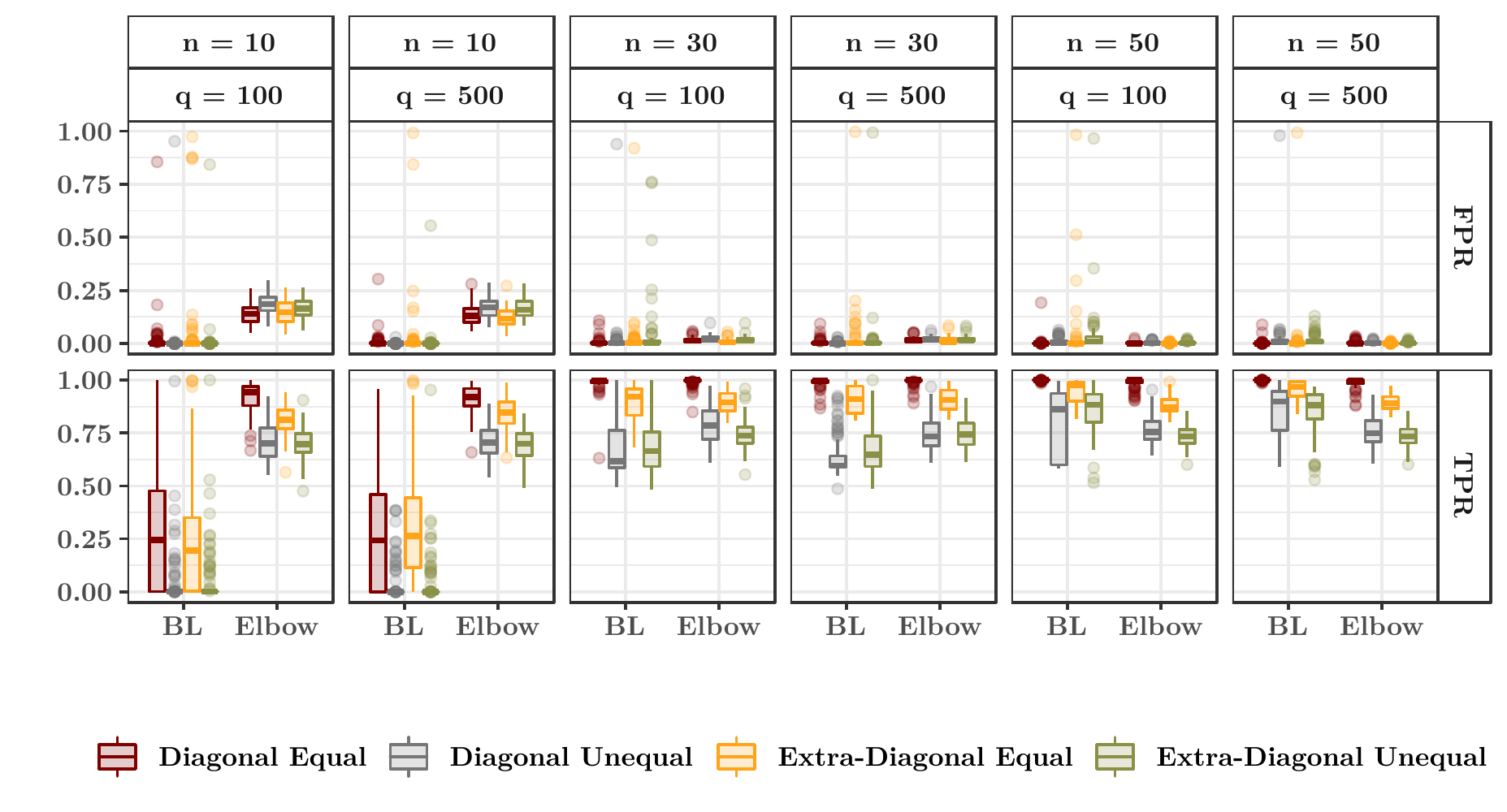}
\caption{Boxplots comparing the TPR (True Positive Rate) and the FPR (False positive Rate) of the two methodologies proposed to select the parameter $\lambda$ 
from 100 replications in the different scenarii.\label{fig:nb_nn0}}
\end{center}
\end{figure}

For the four scenarios, the performance of the two approaches: \textsf{BL} and \textsf{Elbow} described in Section \ref{sec:choix_param} for choosing $\lambda$ 
and hence the number of non null values in $\widetilde{\Sigma}(\lambda)$ 
is illustrated in Figure \ref{fig:nb_nn0}. This figure displays the True Positive Rate (TPR) and the False Positive Rate (FPR) of the methodologies from 100 replications of the matrix $\boldsymbol{E}$
for the four different types of $\boldsymbol{\Sigma}$ and for different values of $n$ and $q$.

\begin{figure}[!h]
\begin{center}
\includegraphics[width=\textwidth]{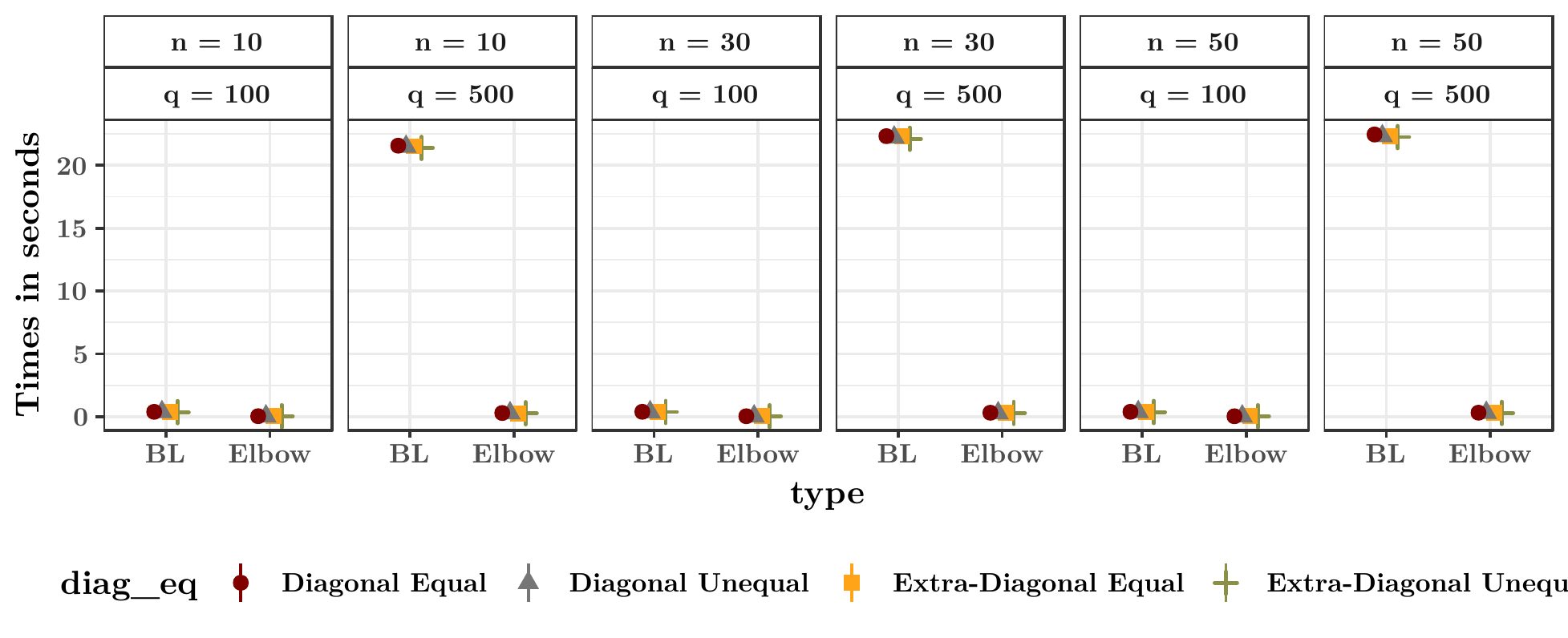}
\caption{Computational times of \textsf{Elbow} and \textsf{BL} criteria. \label{fig:time_nn0}}
\end{center}
\end{figure}

We can see from this figure that the performance of \textsf{Elbow} is on a par with the one of \textsf{BL} except for the case where $n=10$
for which the performance of \textsf{Elbow} is slightly better in terms of True Positive Rate. Moreover, in terms of computational time, the performance 
of \textsf{Elbow} is much better, see Figure \ref{fig:time_nn0}.
 
\subsection{Comparison with other methodologies}\label{sec:comp}

The goal of this section is to compare the statistical performance of our approach with other methodologies.

Since our goal is to estimate a covariance matrix containing blocks, we shall compare our approach with clustering techniques. Once the groups or blocks have be obtained,
$\boldsymbol{\Sigma}$ is estimated by assuming that the corresponding matrix estimator is block-wise constant except for the diagonal blocks for which the diagonal entries
are equal to 1 and the extra-diagonal terms are assumed to be equal. This gives a great advantage to these methodologies
in the \textbf{Diagonal-Equal} and in the \textbf{Extra-Diagonal-Equal} scenarii. More precisely, let  $\rho_{i,j}$ denote the value of the entries in the 
block having its rows corresponding to Group (or Cluster) $i$ and its columns to Group (or Cluster) $j$.
Then, for a given clustering $C$:
\begin{equation}\label{eq:clust2Sig}
\rho_{i,j}= \left\{ \begin{tabular}{lcl}
$\frac{1}{\#C(i)\#C(j)}\displaystyle\sum_{k \in C(i), \ell \in C(j)}R_{k,\ell}$, & if $ C(i) \neq C(j)$ \\
  & \\
 $\frac{1}{\#C(i)(\#C(i)-1)}\displaystyle\sum_{k \in C(i), \ell \in C(i), k\neq \ell}R_{k,\ell}$, & if $ C(i) = C(j)$
 \end{tabular} \right.,
 \end{equation}
 where $C(i)$ denotes the cluster $i$, $\#C(i)$ denotes the number of elements in the cluster $C(i)$ and $R_{k,\ell}$ is the $(k,\ell)$ entry  of the matrix $\boldsymbol{R}$ 
defined in Equation (\ref{eq:mat_corr}).

For the matrices $\boldsymbol{\Sigma}$ corresponding to the four scenarios previously described, we shall compare the statistical performance of the following methods:
\begin{itemize}
\item \textbf{empirical} which estimates $\boldsymbol{\Sigma}$ by $\boldsymbol{R}$ defined in (\ref{eq:mat_corr}),
\item \textbf{blocks} which estimates $\boldsymbol{\Sigma}$ using the methodology described in this article with the criteria \textsf{PA} and \textsf{BL} for choosing $r$ and $\lambda$, respectively,
\item \textbf{blocks\_fast} which estimates $\boldsymbol{\Sigma}$ using the methodology described in this article with the criteria \textsf{Cattell} and \textsf{Elbow} for choosing $r$ and $\lambda$, respectively,
\item \textbf{blocks\_real} which estimates $\boldsymbol{\Sigma}$ using the methodology described in this article when $r$ and the number of non null values are assumed to be known which gives
access to the best value of $\lambda$,
\item \textbf{hclust} which estimates $\boldsymbol{\Sigma}$ by determining clusters using a hierarchical clustering with the ``complete'' agglomeration method
described in \cite{hastie_tibshirani_book} and then uses Equation (\ref{eq:clust2Sig}) to estimate $\boldsymbol{\Sigma}$,
\item \textbf{Specc} which estimates $\boldsymbol{\Sigma}$ by determining clusters using spectral clustering described in \cite{vonLuxburg2007} and estimates $\boldsymbol{\Sigma}$ with Equation (\ref{eq:clust2Sig}),
\item \textbf{kmeans} which estimates $\boldsymbol{\Sigma}$ by determining clusters from a $k$-means clustering approach described in \cite{hastie_tibshirani_book} and then uses Equation (\ref{eq:clust2Sig}) to estimate $\boldsymbol{\Sigma}$.
\end{itemize}

In order to improve the performance of the clustering approaches: \textbf{hclust}, \textbf{Specc} and \textbf{kmeans}, the real number of clusters has been provided to these methods.
The performance of the different approaches is assessed using the Frobenius norm of the difference between $\boldsymbol{\Sigma}$ and its estimator.

\begin{figure}
\includegraphics[width=\textwidth]{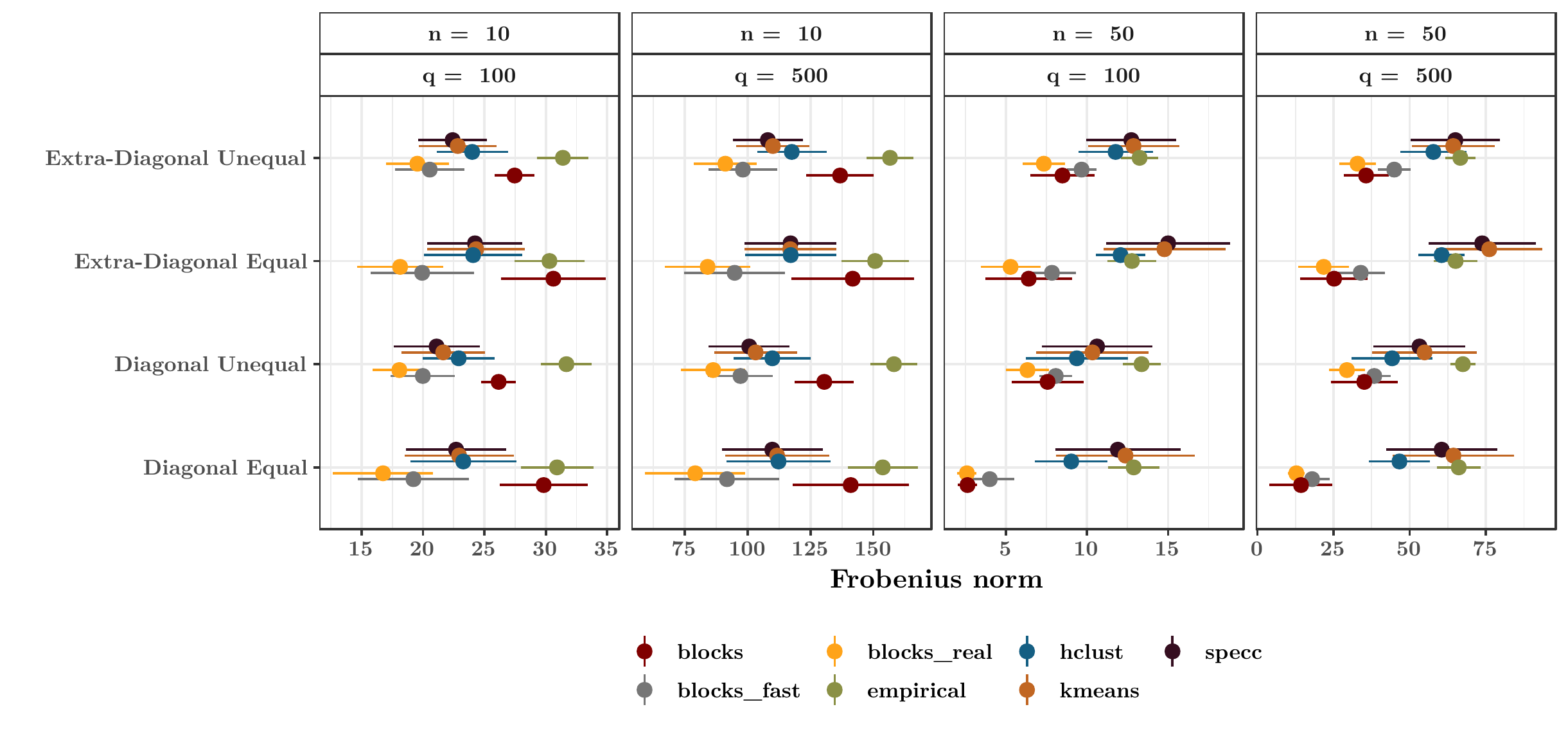}
\caption{Comparison of the Frobenius norm of $\boldsymbol{\Sigma}-\widehat{\boldsymbol{\Sigma}}$ for different estimators $\widehat{\boldsymbol{\Sigma}}$ of $\boldsymbol{\Sigma}$ 
and for different $\boldsymbol{\Sigma}$.\label{fig:comp}}
\end{figure}

Figure \ref{fig:comp} displays the mean and standard deviations of the Frobenius norm of the difference between $\boldsymbol{\Sigma}$ and its estimator 
for different values of $n$ and $q$ in the four different cases:
\textbf{Diagonal-Equal}, \textbf{Diagonal-Unequal}, \textbf{Extra-Diagonal-Equal} and \textbf{Extra-Diagonal-Unequal}.
We can see from this figure that in the case where $n=10$, the performance of \textbf{blocks\_fast} is on a par with the one of \textbf{blocks\_real} and is better than 
the one of \textbf{blocks}. In the case where $n=50$, the performance of \textbf{blocks} is slightly better than the one of \textbf{blocks\_fast} and is similar to the one of \textbf{blocks\_real}.
Moreover, in all cases, either \textbf{blocks\_fast} or \textbf{blocks} outperforms the other approaches.

\begin{figure}[!h]
\begin{center}
\includegraphics[width=0.7\textwidth]{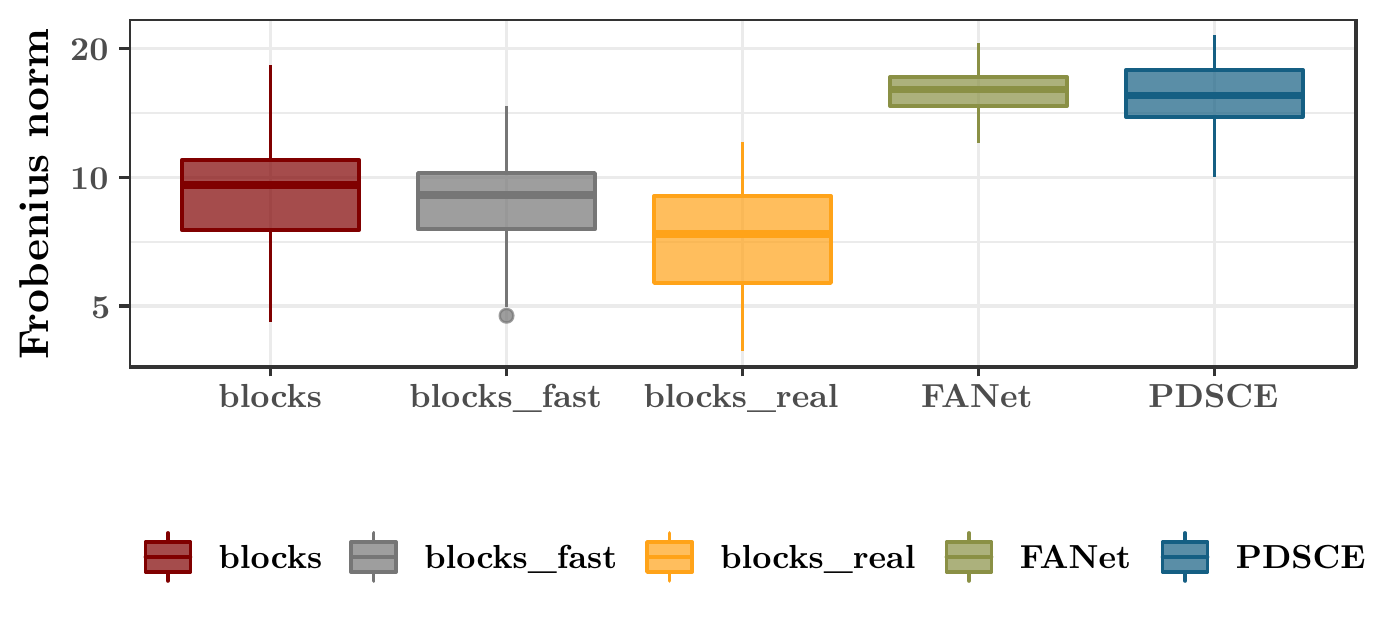}
\caption{Comparison of the Frobenius norm of $\widehat{\boldsymbol{\Sigma}}-\boldsymbol{\Sigma}$
in the \textbf{Extra-Diagonal-Equal} case for $n$ = 30 and $q$ = 100.\label{fig:comp_pdsce}}
\end{center}
\end{figure}

Then, the estimators of $\boldsymbol{\Sigma}$ derived from \textbf{blocks}, \textbf{blocks\_fast} and \textbf{blocks\_real} were compared to the \textbf{PDSCE} estimator proposed by 
\cite{rothman_2012} and implemented in the R package \textbf{PDSCE} and to the  estimator proposed by \cite{blum} and implemented in the \textbf{FANet} package \cite{FANet}. Since the computational
burden of \textbf{PDSCE} is high for large values of $q$, we limit ourselves to the \textbf{Extra-Diagonal-Equal} case when $n=30$ and $q=100$ for the comparison. 
Figure \ref{fig:comp_pdsce} displays the results. We can see from this figure that \textbf{blocks}, \textbf{blocks\_fast} and \textbf{blocks\_real} provide better results 
than \textbf{PDSCE} and \textbf{FANet}. However, it has to be noticed that
\textbf{PDSCE}
is not designed for dealing with block structured covariance matrices but just for providing sparse estimators of large covariance matrices. 

\subsection{Columns permutation}\label{sec:ord}

In practice, it may occur that the columns of $\boldsymbol{E}$ consisting of the rows $\boldsymbol{E}_1,\boldsymbol{E}_2,\dots,\boldsymbol{E}_n$ are not ordered in a way which makes blocks appear
in the matrix $\boldsymbol{\Sigma}$. To address this issue, we propose to perform a hierarchical clustering on $\boldsymbol{E}$ beforehand and use the obtained permutation of the observations
which guarantees that a cluster plot using this ordering will not have crossings of the branches. Let us denote $\boldsymbol{E}_{ord}$ the matrix $\boldsymbol{E}$ in which the columns have been permuted
according to this ordering and $\boldsymbol{\Sigma}_{ord}$ the covariance matrix of each row of $\boldsymbol{E}_{ord}$. Then, we apply our methodology to $\boldsymbol{E}_{ord}$ which should provide
an efficient estimator of $\boldsymbol{\Sigma}_{ord}$. In order to get an estimator of $\boldsymbol{\Sigma}$ the columns and rows are permuted according to the ordering coming from the 
hierarchical clustering. 

\begin{figure}
\includegraphics[width=\textwidth]{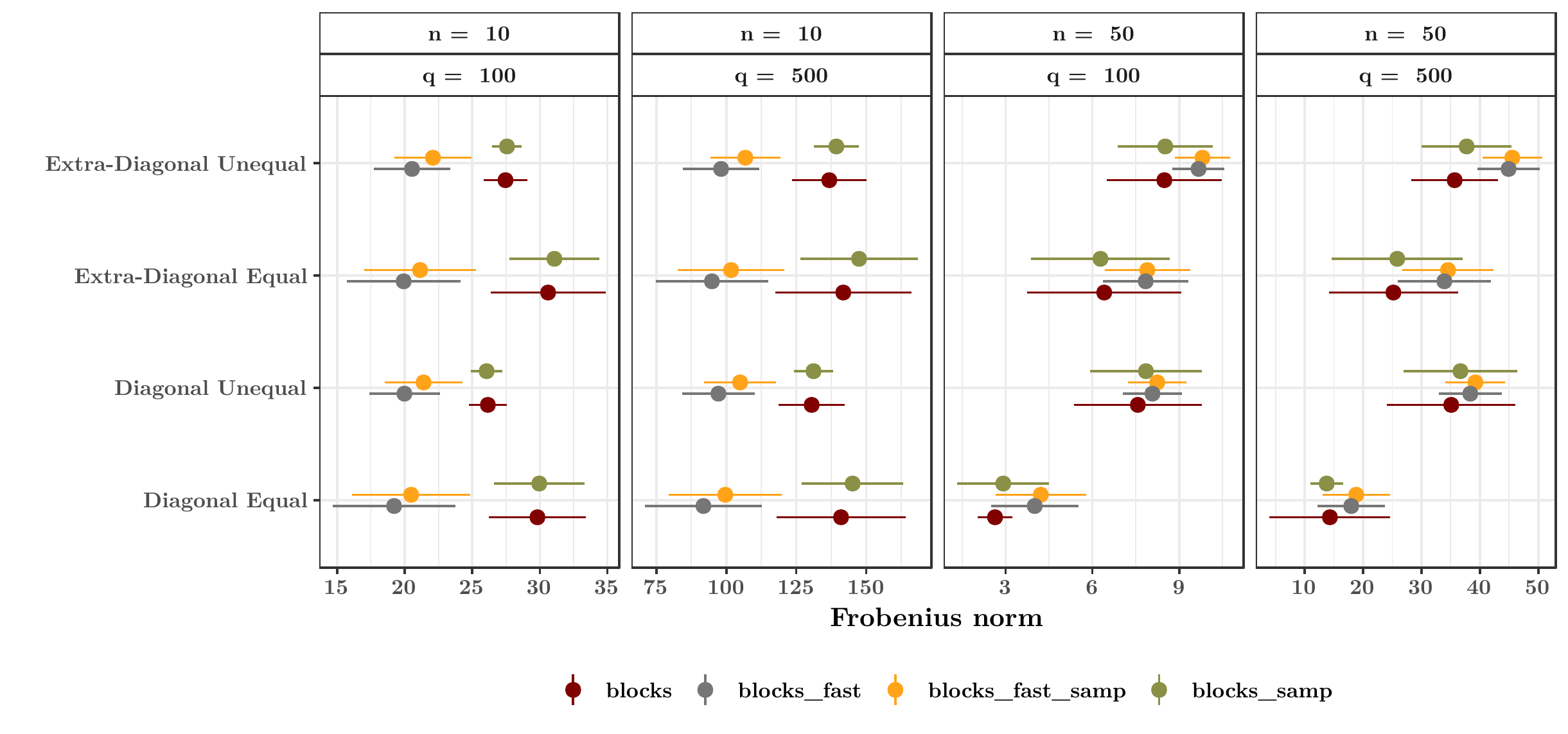}
\caption{Comparison of the Frobenius norm of $\boldsymbol{\Sigma}-\widehat{\boldsymbol{\Sigma}}$, and  $\boldsymbol{\Sigma}_{perm}-\widehat{\boldsymbol{\Sigma}}_{perm}$. \label{fig:comp_samp}}
\end{figure}

To assess the corresponding loss of performance, we generated for each matrix $\boldsymbol{E}$ used for making Figure \ref{fig:comp} a matrix $\boldsymbol{E}_{perm}$ in which the columns
of $\boldsymbol{E}$ were randomly permuted. The associated covariance matrix is denoted $\boldsymbol{\Sigma}_{perm}$. 
Then, we applied the methodology described in the previous paragraph denoted \textbf{blocks\_samp} and \textbf{blocks\_fast\_samp} in Figure \ref{fig:comp_samp} 
thus providing $\widehat{\boldsymbol{\Sigma}}_{perm}$.
The performance of this new methodology was compared to the methodology that we proposed in the previous sections (denoted \textbf{blocks} and \textbf{blocks\_fast} in Figure \ref{fig:comp_samp}) 
when the columns of $\boldsymbol{E}$ were not permuted. The results are displayed in Figure \ref{fig:comp_samp}.
We can see from this figure that the performance of our approach does not seem to be altered by the permutation of the columns.

\subsection{Numerical performance}

\begin{figure}[htp]
\centering
\includegraphics[scale=0.7]{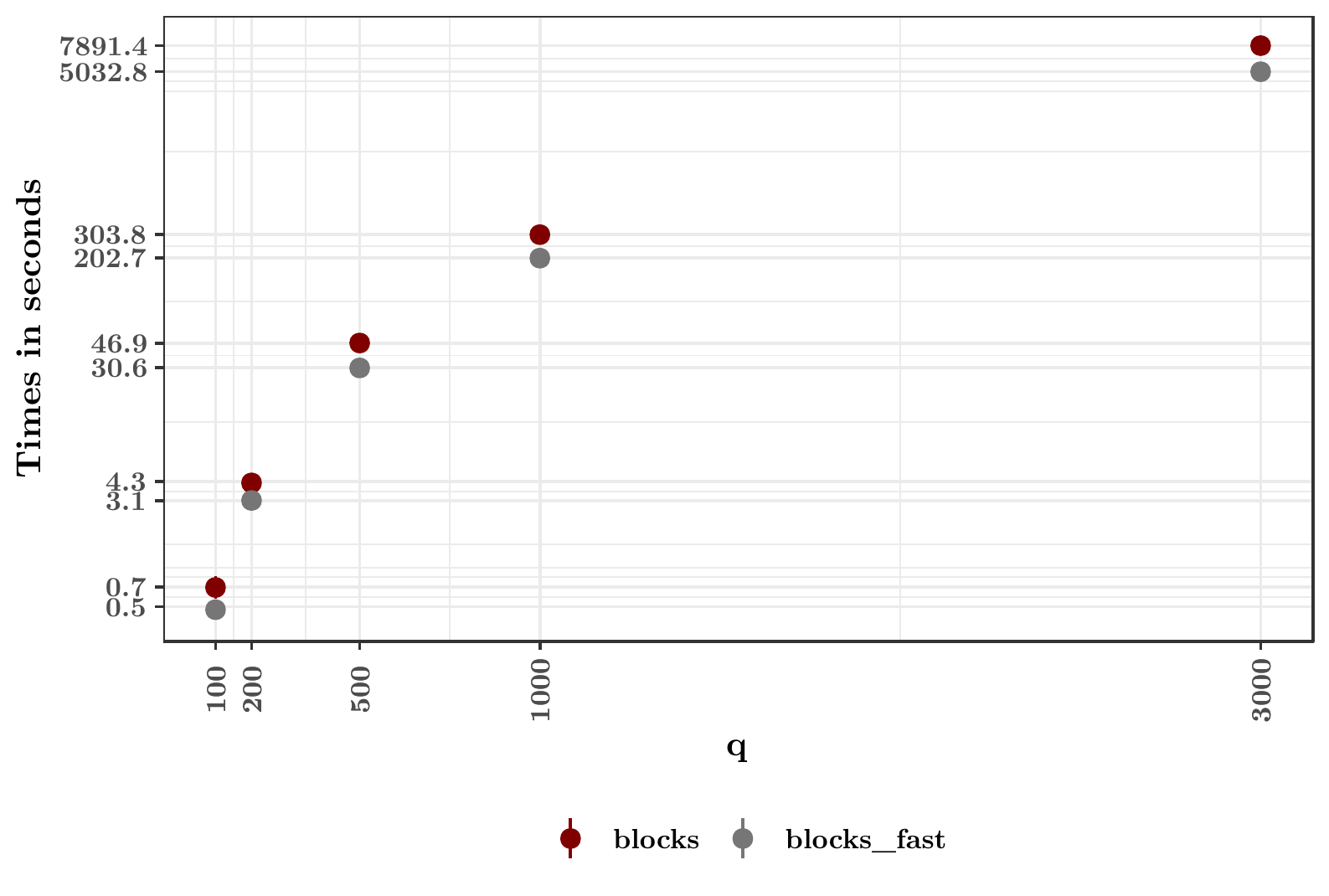}
\caption{Times in seconds to perform our methodology in the \textbf{Extra-Diagonal Unequal} case. \label{fig:tps}}
\end{figure}

Figure \ref{fig:tps} displays the computational times for estimating $\boldsymbol{\Sigma}$ with the methods \textbf{blocks} and \textbf{blocks\_fast} 
for different values of $q$ ranging from 100 to 3000 and $n=30$. The timings were obtained on a workstation with
16 GB of RAM and Intel Core i7 (3.66GHz) CPU. Our methodology is implemented in the R package \texttt{BlockCov} which uses the R language (R Core
Team, 2017) and relies on the R package \texttt{Matrix}. We can see from this figure that it takes around 3 minutes to estimate a $1000\times 1000$ correlation matrix.

\subsection{Choice of the threshold $t$ for estimating $\boldsymbol{\Sigma}^{-1/2}$}\label{sec:choice_t}

\begin{figure}[htp]
\centering
\includegraphics[width=\textwidth]{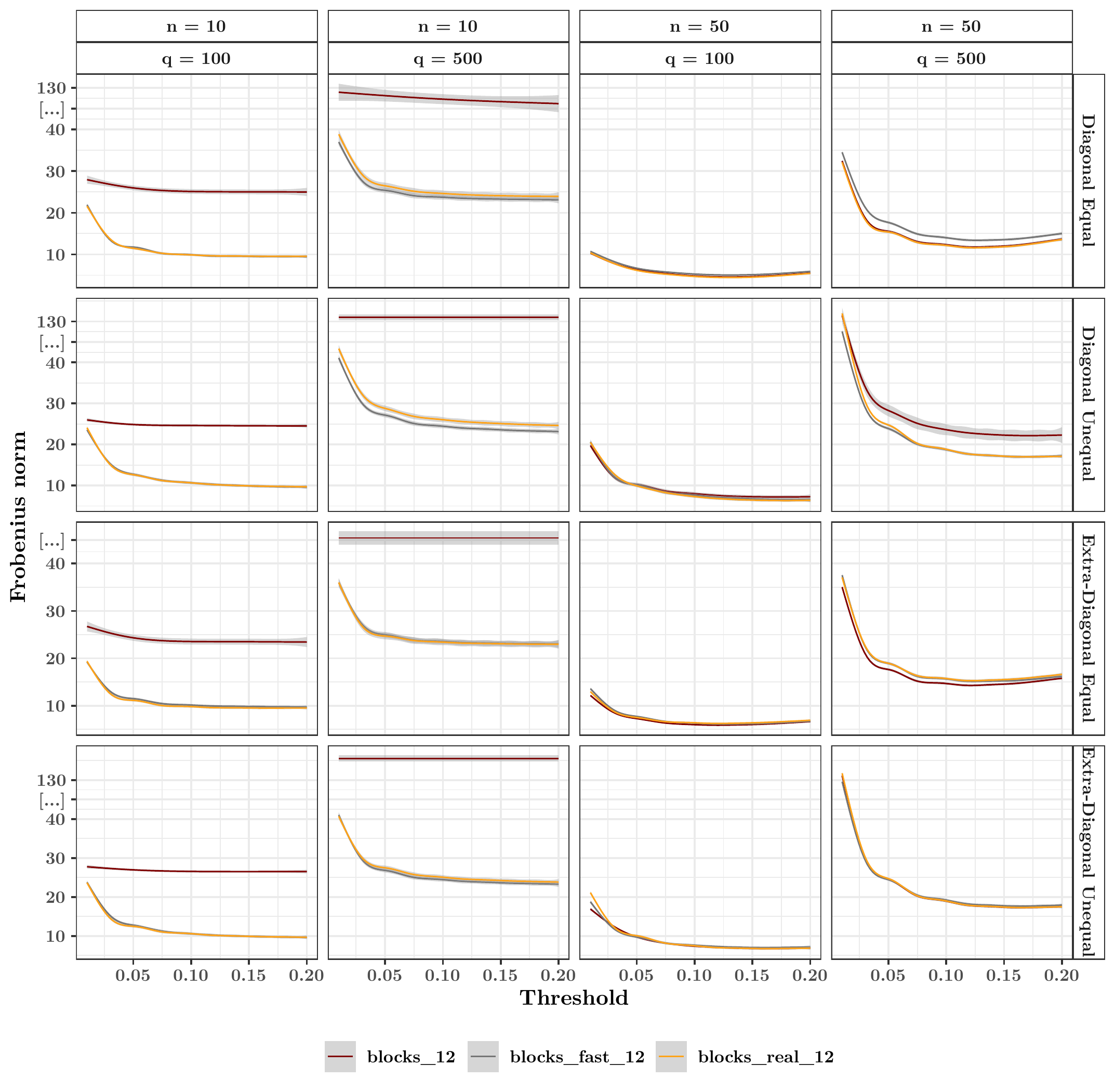}
\caption{Frobenius norm of $\widehat{\boldsymbol{\Sigma}}_t^{-1/2}\boldsymbol{\Sigma}\widehat{\boldsymbol{\Sigma}}_t^{-1/2}-\textrm{Id}_q$, 
where $\widehat{\boldsymbol{\Sigma}}^{-1/2}_t$ is computed for different 
thresholds $t$.\label{fig:tresh}}
\end{figure}

Since we are interested in assessing the ability of $\widehat{\boldsymbol{\Sigma}}^{-1/2}_t$ defined in (\ref{eq:Sigma-1/2_t}) 
to remove the dependence that may exist between the columns of $\boldsymbol{E}$, we shall consider
the Frobenius norm of $\widehat{\boldsymbol{\Sigma}}^{-1/2}_t\boldsymbol{\Sigma}\widehat{\boldsymbol{\Sigma}}^{-1/2}_t-\textrm{Id}_q$
which should be close to zero, where $\textrm{Id}_q$ denotes the identity matrix of $\mathbb{R}^q$.
 Figure \ref{fig:tresh} displays the Frobenius norm of $\widehat{\boldsymbol{\Sigma}}^{-1/2}_t\boldsymbol{\Sigma}\widehat{\boldsymbol{\Sigma}}^{-1/2}_t-\textrm{Id}_q$ for different threshold $t$. A threshold of 0.1 seems to provide a small error in terms of Frobenius norm. Hence, in the following, $t$ 
will be equal to 0.1 and $\widehat{\boldsymbol{\Sigma}}^{-1/2}_{0.1}$ will be referred as $\widehat{\boldsymbol{\Sigma}}^{-1/2}$.

\begin{figure}
\includegraphics[width=\textwidth]{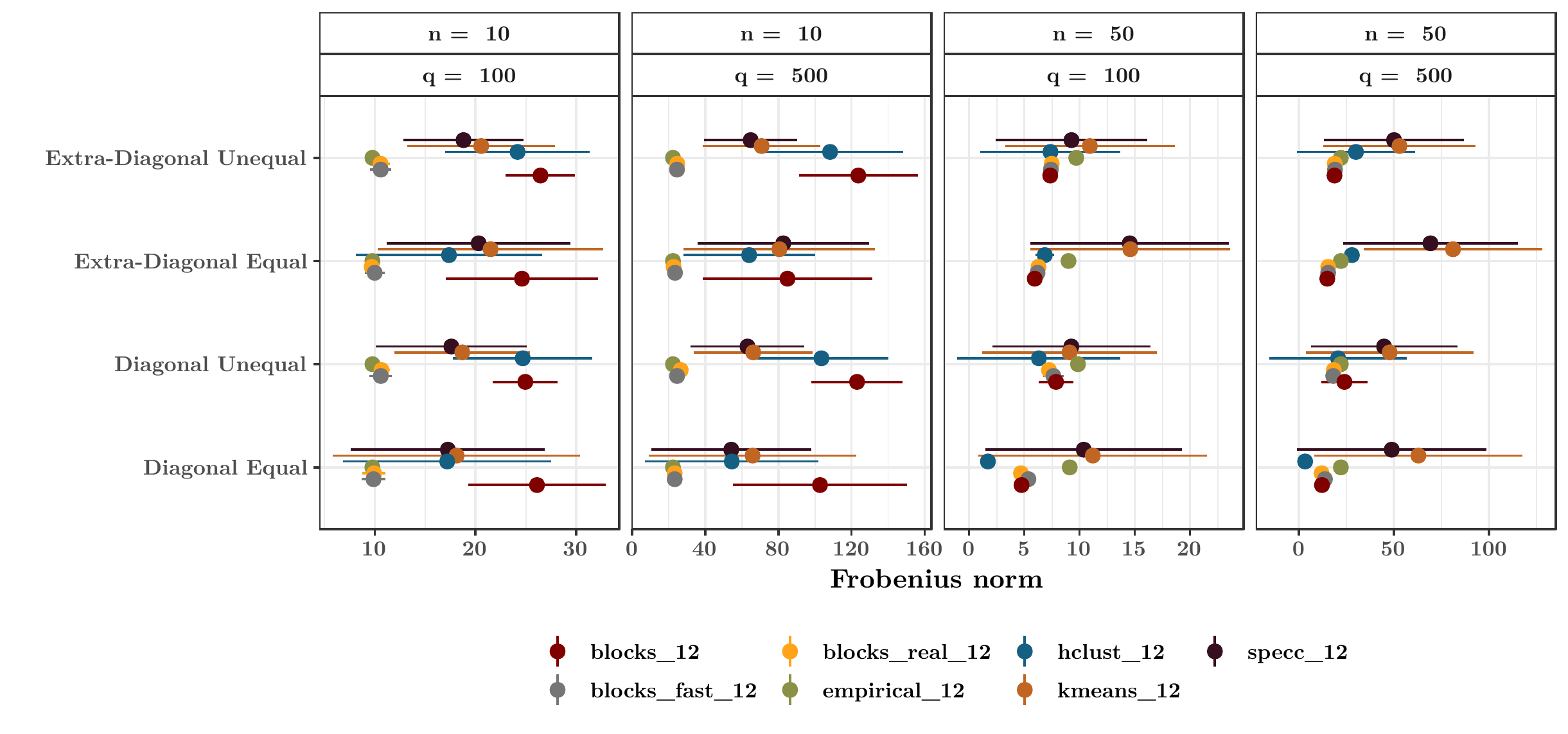}
\caption{Comparison of the Frobenius norm of the error $\widehat{\boldsymbol{\Sigma}}^{-1/2}\boldsymbol{\Sigma}\widehat{\boldsymbol{\Sigma}}^{-1/2}-\textrm{Id}_q$, 
for different estimators $\widehat{\boldsymbol{\Sigma}}$ of $\boldsymbol{\Sigma}$.
\label{fig:comp_12}}
\end{figure}

This technique was applied to all of the estimators of $\boldsymbol{\Sigma}$ discussed in Section \ref{sec:comp} to get different estimators of $\boldsymbol{\Sigma}^{-1/2}$.
The Frobenius norm of the error $\widehat{\boldsymbol{\Sigma}}^{-1/2}\boldsymbol{\Sigma}\widehat{\boldsymbol{\Sigma}}^{-1/2} - \textrm{Id}_q$ is used to 
compare the different estimators obtained by considering the different estimators of $\boldsymbol{\Sigma}$.
The results are displayed in Figure \ref{fig:comp_12}. We observe from this figure that in the case where $n =10$ the estimators of $\boldsymbol{\Sigma}^{-1/2}$ derived from the \textbf{empirical}, the   \textbf{blocks\_fast} and the \textbf{blocks\_real} estimators of $\boldsymbol{\Sigma}$ perform similarly and seem to be more adapted than the others to remove the dependence among the columns of $\boldsymbol{E}$. 
However, when $n=50$, the behavior is completely different.  Firstly, in the \textbf{Diagonal-Equal} case, the 
estimator of $\boldsymbol{\Sigma}^{-1/2}$ derived from the \textbf{hclust} estimator of $\boldsymbol{\Sigma}$ seems to perform better than the others. Secondly, in the \textbf{Diagonal-Unequal} case,
 the estimator derived from \textbf{blocks}, \textbf{blocks\_fast} and \textbf{blocks\_real} perform similarly than the one obtained from  \textbf{hclust}. 
Thirdly, in the \textbf{Extra-Diagonal} case, the estimators derived from \textbf{blocks}, \textbf{blocks\_fast} and \textbf{blocks\_real} methodology perform
better than the other estimators.  

\begin{figure}[!h]
\begin{center}
\includegraphics[width=0.7\textwidth]{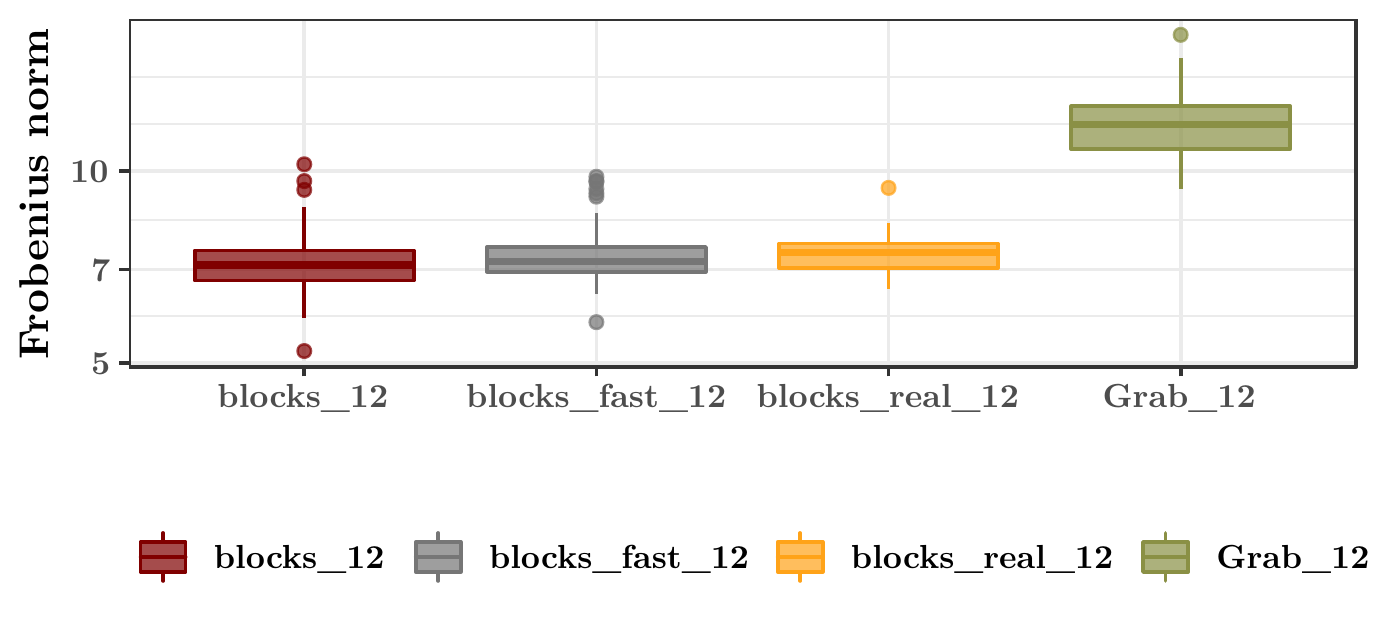}
\caption{Comparison of the Frobenius norm of $\widehat{\boldsymbol{\Sigma}}^{-1/2}\boldsymbol{\Sigma}\widehat{\boldsymbol{\Sigma}}^{-1/2}-\textrm{Id}_q$
in the \textbf{Extra-Diagonal-Equal} case.\label{fig:comp_grab}}
\end{center}
\end{figure}

Then, the estimators of $\boldsymbol{\Sigma}^{-1/2}$ derived from \textbf{blocks}, \textbf{blocks\_fast} and \textbf{blocks\_real} 
were compared to the \textbf{GRAB} estimator proposed by \cite{GRAB}. Since the computational
burden of \textbf{GRAB} is high for large values of $q$, we limit ourselves to the \textbf{Extra-Diagonal-Equal} case when $n=30$ and $q=100$ for the comparison. 
Figure \ref{fig:comp_grab} displays the results. We can see that \textbf{blocks} and \textbf{blocks\_real} provide better results than \textbf{GRAB}. However, it has to be noticed that
the latter approach depends on a lot of parameters that were difficult to choose, thus we used the default ones.

\subsection{Use of  $\boldsymbol{\Sigma}^{-1/2}$ to remove the dependence in multivariate linear models}\label{sec:choice_t}

Eventually, we assess the performance of the BlockCov methodology to remove the dependence in the columns of an observation matrix in order to be used for variable selection in the multivariate linear model as it is performed in the \texttt{MultiVarSel} R package: 
\begin{equation}\label{eq:mod_lin_chap_4}
\bY = \bX\bB+ \bE,
\end{equation}
where $\bY$ is a $n \times q$ response matrix, $\bX$ is a $n  \times p$ design matrix, $\bB$ is a coefficients matrix and $\bE$ is an error matrix.
Here, $\boldsymbol{E}_1, \boldsymbol{E}_2,\cdots,\boldsymbol{E}_n$ are  $n$ zero-mean i.i.d. $q$-dimensional 
Gaussian random vectors having a covariance matrix 
$\boldsymbol{\Sigma}$.
To achieve this goal, we generate observations
$\boldsymbol{Y}$ according to this  multivariate linear model.
We choose $q =100$, 
$p=3$, $n=30$ and $bX$ is the design matrix of a one-way ANOVA model.
We compared our methodology with the one proposed by \cite{perthame2016stability} and implemented in the \texttt{FADA} R package \cite{FADA}.
We shall investigate the effect of the sparsity of $\bB$ and of the signal
to noise ratio (SNR) for the four scenarii defining $\boldsymbol{\Sigma}$ on the selection of the non null values of $\bB$ in (\ref{eq:mod_lin_chap_4}).
Different signal to noise ratios are obtained by multiplying $\bB$ in (\ref{eq:mod_lin_chap_4})  by a
coefficient $\kappa$.

\begin{figure}[!h]
\begin{center}
\begin{tabular}{cl}
\includegraphics[width=0.5\textwidth]{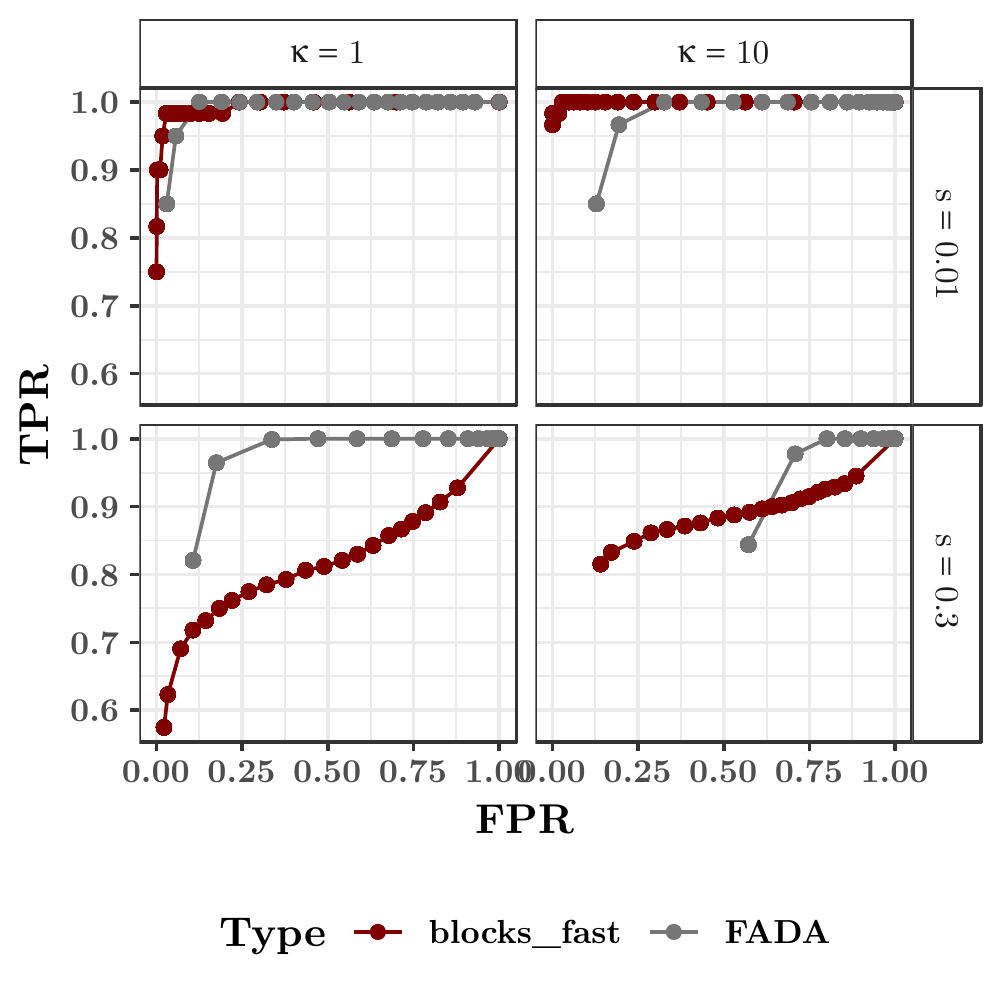}&
\includegraphics[width=0.5\textwidth]{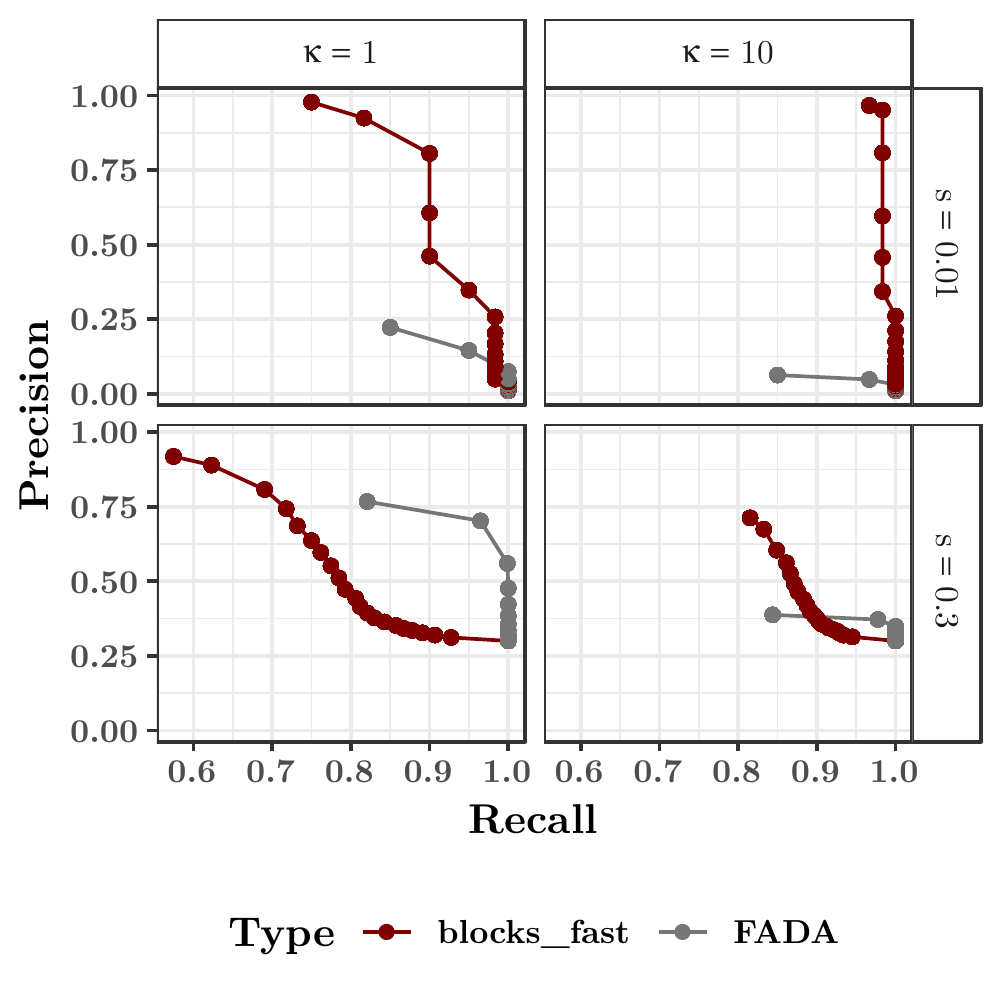}
\end{tabular}
\caption{Means of the ROC curves (left) and Precision Recall curves (right) obtained from 100 replications comparing the variables selected by the MultiVarSel strategy using either $\boldsymbol{\Sigma}^{-1/2}$ obtained by BlockCov to remove the dependence or  the methodology proposed by FADA methodology. $\kappa$ is linked to the signal to noise ratio and $s$ denotes the sparsity levels \textit{i.e} the fraction of non-zero elements in $\bB$.\label{fig:comp_fada}}
\end{center}
\end{figure}

Since the results are barely influenced by the scenario chosen for $\boldsymbol{\Sigma}$, only the \textbf{Extradiagonal-Equal} case is displayed in 
Figure~\ref{fig:comp_fada}, the other scenarii are available in Annexe~\ref{FADA}. 
We can see from this figure that when the signal to noise ratio is low and the value of $s$ is high, meaning that there is a lot of non-zero values, 
the FADA methodology performs better than the BlockCov methodology. 
Nevertheless, in the three other cases the performance of BlockCov is either better or on a par with the one of FADA methodology.

\section{Application to ``multi-omic'' approaches to study seed quality}\label{sec:real_data}

Climate change could lead to major crop failures in world. In the present study, we addressed the impact of mother plant environment on seed composition. Indeed, seed quality is of paramount ecological and agronomical importance. They are the most efficient form of dispersal of flowering plants in the environment. Seeds are remarkably adapted to harsh environmental conditions as long as they are in a quiescent state. Dry mature seeds (so called “orthodox seeds”) are an appropriate resource for preservation of plant genetic diversity in seedbanks. It has been reported that the temperature regime during seed production affects agronomical traits such as seed germination potential, see \cite{huang_2014},\cite{macgregor_2015} and
\cite{kerdaffrec_2017}. In order to highlight biomarkers of seed quality according to thermal environment of the mother plant, Arabidopsis seeds were produces under three temperature regimes 
(14-16 $^o$C, 18-22 $^o$C or 25-28 $^o$C under a long-day photoperiod). Dry mature seeds were analysed by shotgun proteomic and GC/MS-based metabolomics \cite{durand_2019}. 
The choice to use the model plant, Arabidopsis, was motivated by the colossal effort of the international scientific community for its genome annotation. 
This plant remains at the forefront of modern genetics, genomics, plant modelling and system biology, see \cite{provart_2016}. Arabidopsis provides 
a very useful basis to study gene regulatory networks, and develop modelling and systems biology approaches for translational research towards agricultural applications.

In this section, we apply  our R packages \texttt{BlockCov} and \texttt{MultiVarSel} \cite{packageMultiVarSel} to metabolomic and proteomic data 
to better understand the impact of the temperature on the seed quality.
More precisely, we use the following modeling for our observations:
\begin{equation}\label{mod:metab}
\bY=\bX\bB+\bE,
\end{equation}
where $\bY$ is a $n \times q$ matrix containing the responses of the $q$ metabolites (resp. the $q$ proteins) for the $n$ samples with $n=9$, $q=199$ (resp. $q=724$) 
for the metabolomic (resp. proteomic) dataset, $\bX$ is a $n \times 3$ design matrix of a one-way ANOVA model, 
such that its first (resp. second, resp.  third) column is a vector which is equal to 1 if the corresponding sample grows under low (resp medium, resp. elevated) temperatures and 0 otherwise. 
$\bB$ is a coefficient matrix and $\bE$ is such that its $n$ rows $\boldsymbol{E}_1, \boldsymbol{E}_2,\cdots,\boldsymbol{E}_n$ are $n$ zero-mean i.i.d. $q$-dimensional 
random vectors having a covariance matrix $\boldsymbol{\Sigma}$. We used our R package \texttt{BlockCov} to estimate $\boldsymbol{\Sigma}$ and $\boldsymbol{\Sigma}^{-1/2}$ 
assuming that there exists a latent
block structure in the covariance matrix of the rows of $\boldsymbol{E}$. More precisely, we assume that there exists some groups of metabolites (resp. proteins) 
having the same behavior since they belong
to the same biological process. Then, we plugged this estimator into our R package
\texttt{MultiVarSel} to obtain a sparse estimation of $\bB$. Thanks to this estimator of $\bB$, we could identify the metabolites (resp. proteins) having a higher (resp. lower) concentration when
the temperature is high or low.

\subsection{Results obtained for the metabolomic data}

\begin{figure}[!h]
\begin{center}
\includegraphics[width=0.9\textwidth]{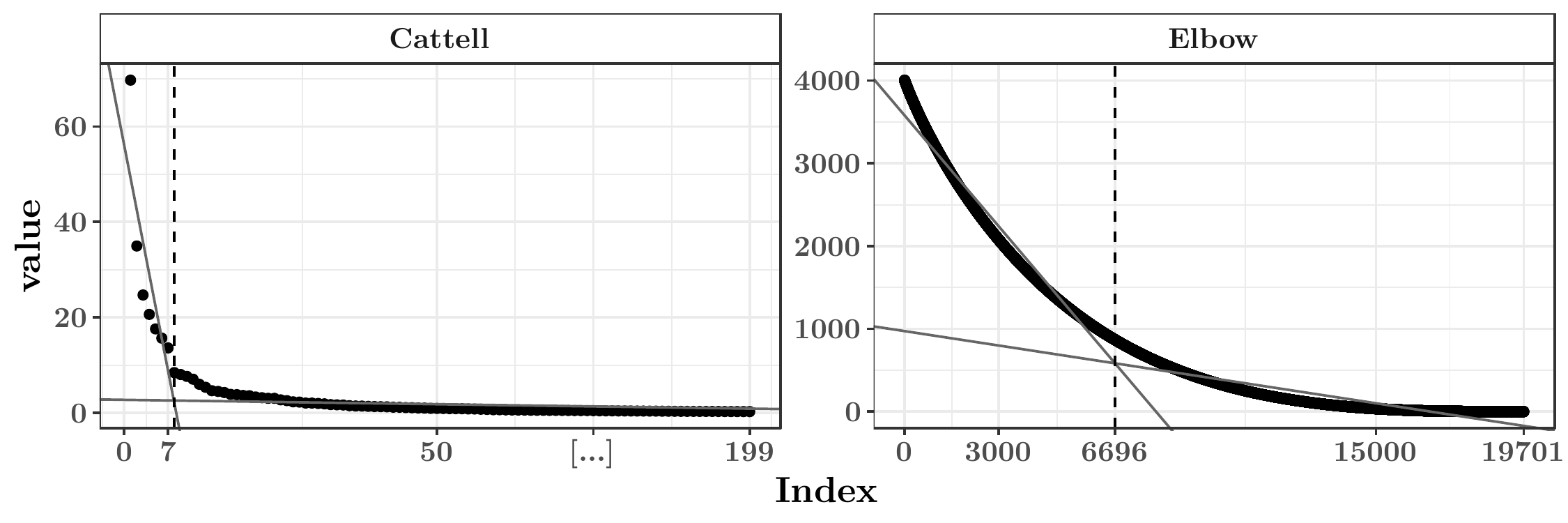}
\caption{Illustration of the \textsf{Cattell} and \textsf{Elbow} criteria.\label{fig:k_metab}}
\end{center}
\end{figure}

We first estimated the matrices $\boldsymbol{\Sigma}$ and $\boldsymbol{\Sigma}^{-1/2}$ associated to $\boldsymbol{E}$ 
defined in Equation (\ref{mod:metab}) by using the methodology developed in this paper, namely the \texttt{BlockCov} package.
By the results of Section \ref{sec:num_exp}, we know that the \textsf{PA} and \textsf{BL} approaches performed poorly when $n =10$. Since here $n=9$, we used
the \textsf{Cattell} and \textsf{Elbow} criteria to choose $r$ and $\lambda$, respectively. 
The results are displayed  in Figure~\ref{fig:k_metab}. The \textsf{Cattell} criterion chooses $r =7$ and the \textsf{Elbow} criterion chooses $\lambda= 0.472$, which implies that among the 19701 coefficients of the correlation matrix only 6696 values are considered as non null values. 

\begin{figure}
\includegraphics[width=0.7\textwidth]{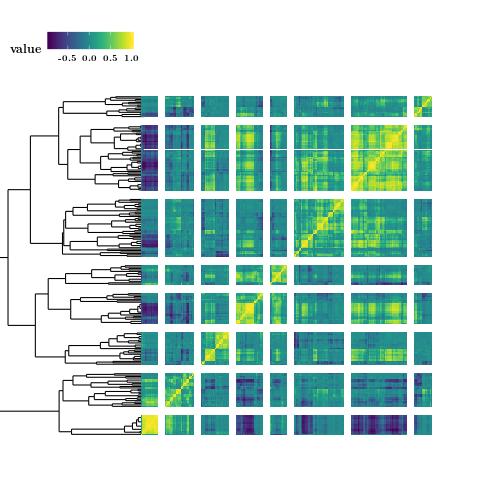}
\caption{Estimator of the correlation matrix  $\boldsymbol{\Sigma}$ of the rows of $\bE$ once the rows and 
the columns have been permuted according to the ordering provided by the hierarchical clustering.\label{fig:cor_metab}}
\end{figure}

The estimation of $\boldsymbol{\Sigma}$ obtained with our methodology is displayed in Figure~\ref{fig:cor_metab} once the rows and 
the columns have been permuted according to the ordering provided by the hierarchical clustering to make visible the latent block structure.

\begin{figure} 
\hspace{-5mm}
\includegraphics[width=0.6\textwidth]{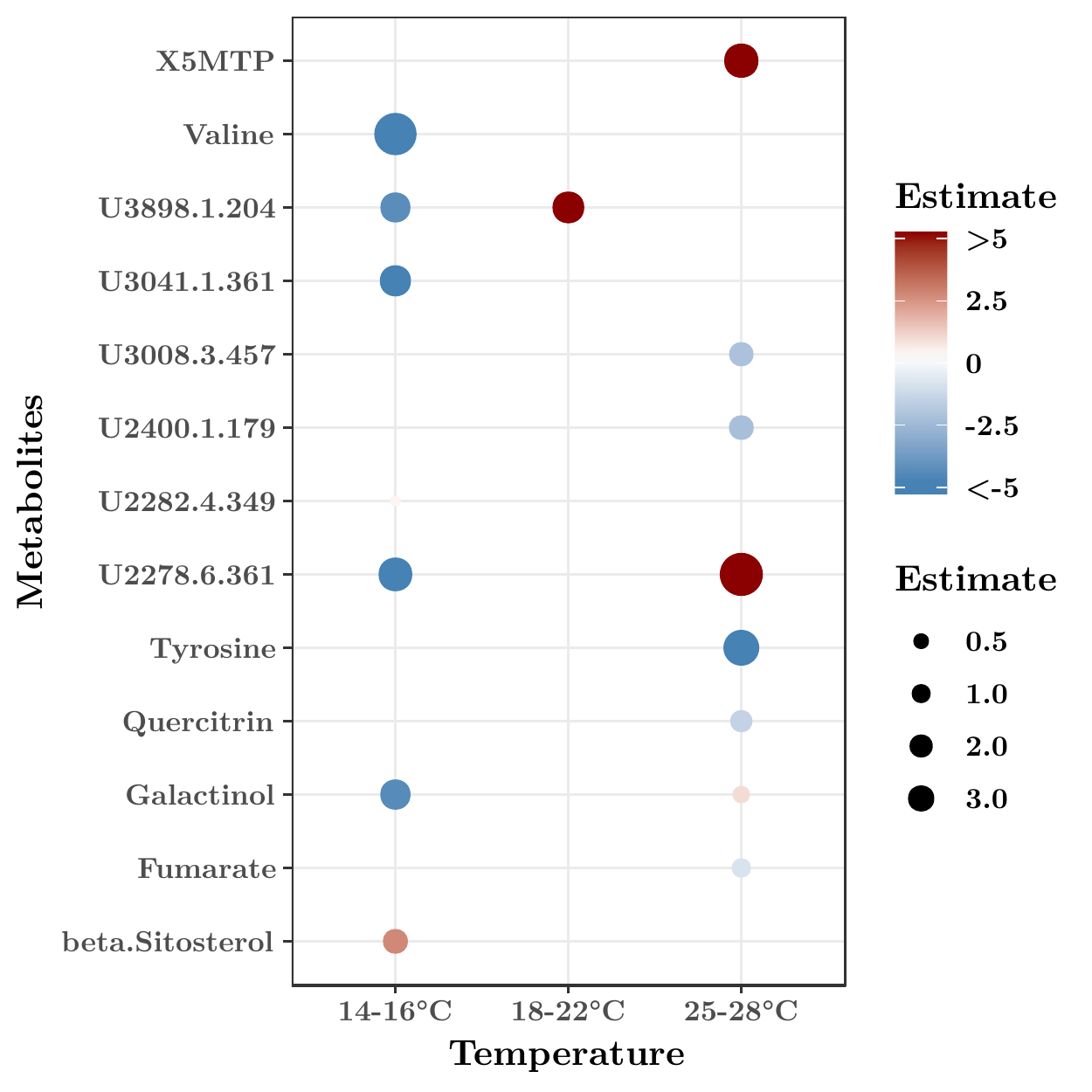}
\caption{Sparse estimator of the coefficients matrix $\bB$ obtained thanks to the package \texttt{MultiVarSel} with a threshold of 0.95. \label{fig:metab_coeff}}
\end{figure}

Using the estimator of $\boldsymbol{\Sigma}^{-1/2}$ provided by the \texttt{BlockCov} package in the R package \texttt{MultiVarSel} provides
the sparse estimator of the matrix $\bB$ defined in Model~\ref{mod:metab} and displayed in Figure \ref{fig:metab_coeff}.
We can see from this figure that for the metabolite X5MTP the coefficient of the matrix $\widehat{\bB}$ is positive when the temperature is high which
means that the production of the metabolite X5MTP is larger in high temperature conditions than in low temperature conditions.

In order to go further in the biological interpretation,  we wanted to better understand the underlying block structure of the estimator of the correlation matrix of the residuals 
based on metabolite abundances $\widehat{\boldsymbol{\Sigma}}$. Thus, we applied a hierarchical clustering with 8 groups to this matrix
in order to split it into blocks. The corresponding dendogram is on the left part of Figure \ref{fig:cor_metab}.
The matrix containing the correlation means within and between the blocks or groups of metabolites is displayed in Figure~\ref{fig:mcormeta}.
The composition of the metabolites groups is available in Appendix~\ref{sec:ap}.

\begin{figure}
\hspace{-5mm}
\includegraphics[width=0.6\textwidth]{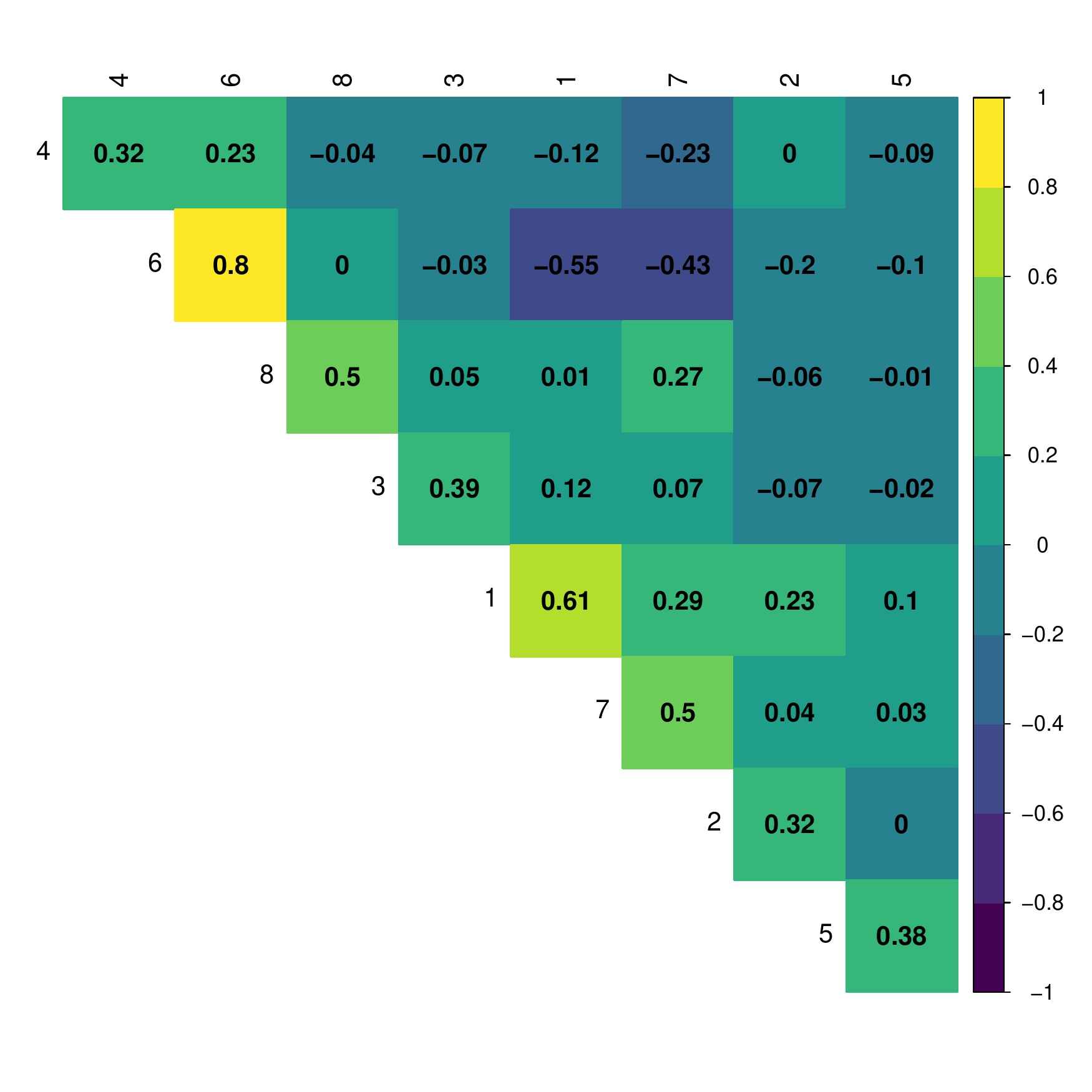}
\caption{Means of the correlations between the groups of metabolites.\label{fig:mcormeta}}
\end{figure}
 
Interestingly, we could observe that X5MTP belongs to Group 6 which displays an high correlations mean equal to 0.8 between the 14 metabolites that make it up. 
At least 6 metabolites of this group belong to the same family,
namely glucosinolates (i.e. X4MTB, 4-methylthiobutyl glucosinolate; X5MTP, 5-methylthiopentyl
glucosinolate; X6MTH, 6-methylthiohexyl glucosinolate; X7MTH, 7-methylthiohexyl glucosinolate;
X8MTO, 8-methylthiooctyl glucosinolate; UGlucosinolate140.1, unidentified glucosinolate).
Glucosinolates (GLS) are specialized metabolites found in Brassicaceae and related families (e.g.
Capparaceae), containing a $\beta$-thioglucose moiety, a sulfonated oxime moiety, and a variable aglycone
side chain derived from a $\alpha$-amino acid. These compounds contribute to the plant's overall defense
mechanism, see \cite{wittstock_2002}. Methylthio-GLS are derivated from methionine.
Methionine is elongated through condensation with acetyl CoA and then, are converted to aldoximes
through the action of individual members of the cytochrome P450 enzymes belonging to CYP79
family, see \cite{field_2004}. The aldoxime undergoes condensation with a sulfur donor, and stepwise
converted to GLS, followed by the side chain modification.
The present results suggest that the
accumulation of methionine-derived glucosinolate family is strongly coordinated in Arabidopsis seed.
Moreover, we can see that they are influenced by the effect of the mother plant thermal environment.

\subsection{Results obtained for the proteomic data}

\begin{figure}
\hspace{-5mm}
\includegraphics[width=0.7\textwidth]{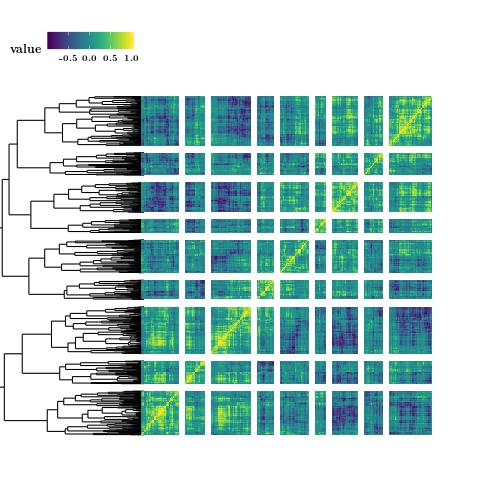}
\caption{Estimator of the correlation matrix of the residuals of the protein accumulation measures once the rows and 
the columns of the residual matrix have been permuted according to the ordering provided by the hierarchical clustering.\label{fig:cor_prot}}
\end{figure}

The same study was conducted on the proteomic data. The estimator of the correlation matrix of the residuals 
based on proteine abundances $\widehat{\boldsymbol{\Sigma}}$ obtained with our methodology is displayed in Figure~\ref{fig:cor_prot} once the rows and 
the columns have been permuted according to the ordering provided by the hierarchical clustering to make visible the latent block structure.
To better understand the underlying block structure of $\widehat{\boldsymbol{\Sigma}}$, we applied a hierarchical clustering with 9 groups to this matrix
in order to split it into blocks. The corresponding dendogram is on the left part of Figure \ref{fig:cor_prot}.

\begin{figure}
\hspace{-5mm}
\includegraphics[width=0.6\textwidth]{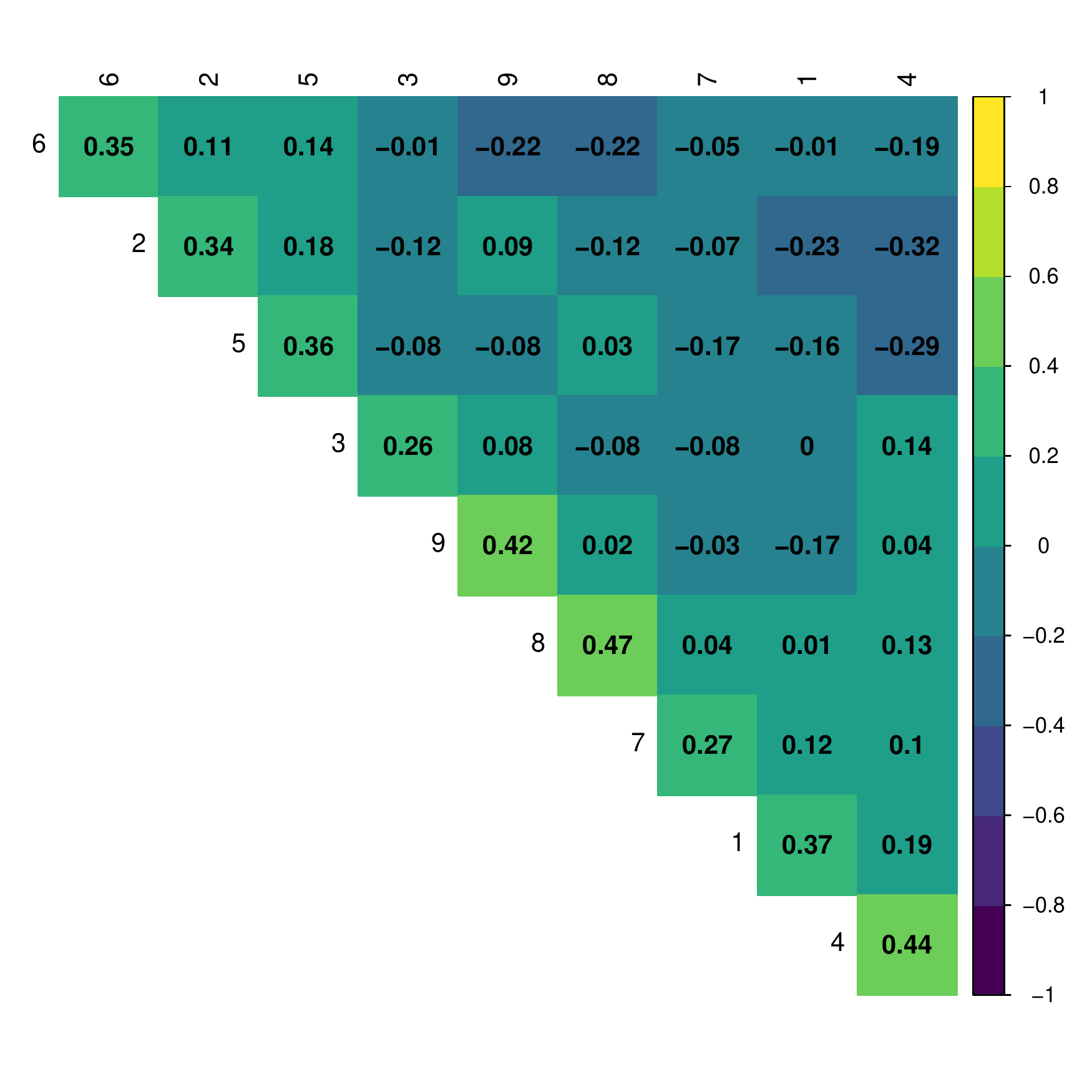}
\caption{Means of the correlations between the groups of proteins.\label{fig:mcorp}}
\end{figure}

The matrix containing the correlation means within and between the blocks or groups of proteins is displayed in Figure~\ref{fig:mcorp}.
We can see from this figure that Group 8 has the highest correlation mean equal to 0.47. It consists of 34 proteins which are given in 
Appendix~\ref{sec:protgroup}.

\begin{figure}
\hspace{-5mm}
\includegraphics[width=0.9\textwidth]{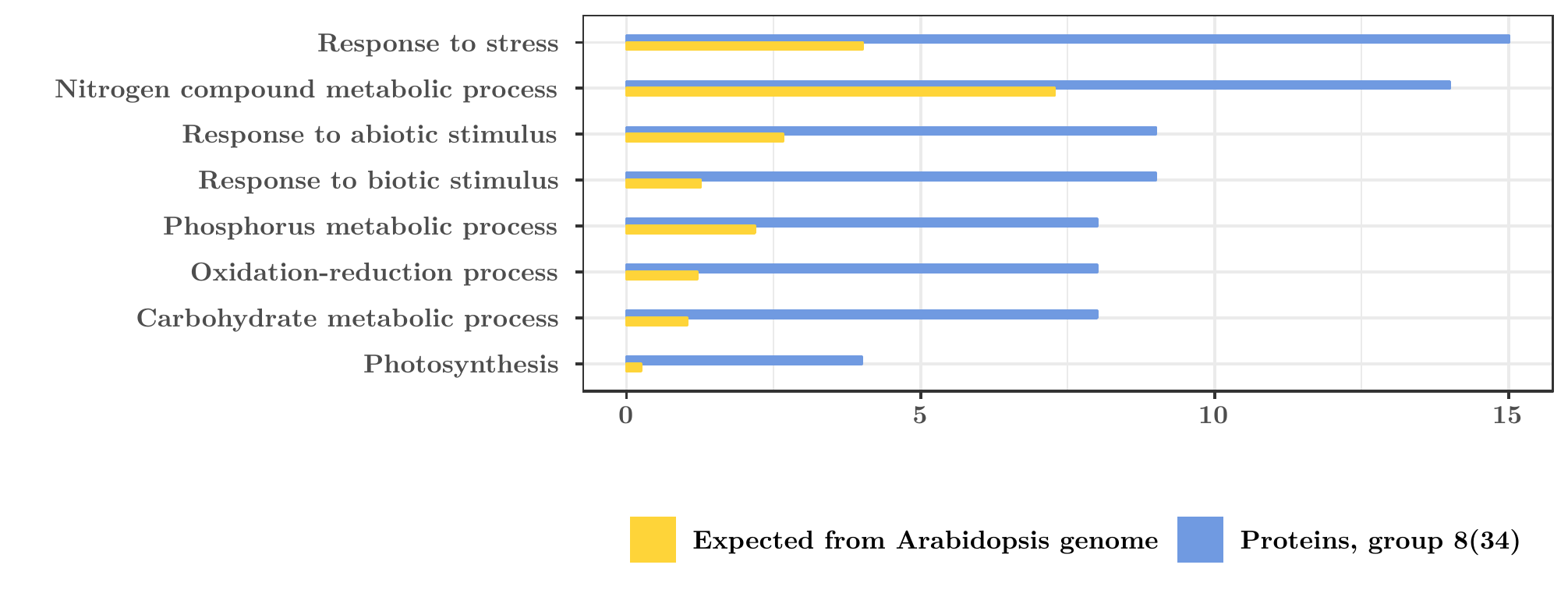}
\caption{Gene ontology (GO) term enrichment analysis of the 34 proteins belonging to Group 8. Data from PANTHER overrepresentation test (http://www.geneontology.org); One uploaded id (i.e. AT5G50370) mapped to two genes. Thus, GO term enrichment was performed on 35 elements. Blue bars: observed proteins in Group 8; 
Orange bars: expected result from the reference Arabidopsis genome.\label{fig:go}}
\end{figure}

A basic gene ontology analysis (http://geneontology.org/) showed that proteins involved in response to stress (biotic
and abiotic), in nitrogen and phosphorus metabolic processes, in photosynthesis and carbohydrate
metabolic process and in oxidation-reduction process are overrepresented in this group, see Figure~\ref{fig:go}. Thus,
the correlation estimated within Group 8 seems to reflect a functional coherence of
the proteins of this group.

\begin{figure}
\hspace{-5mm}
\includegraphics[width=0.6\textwidth]{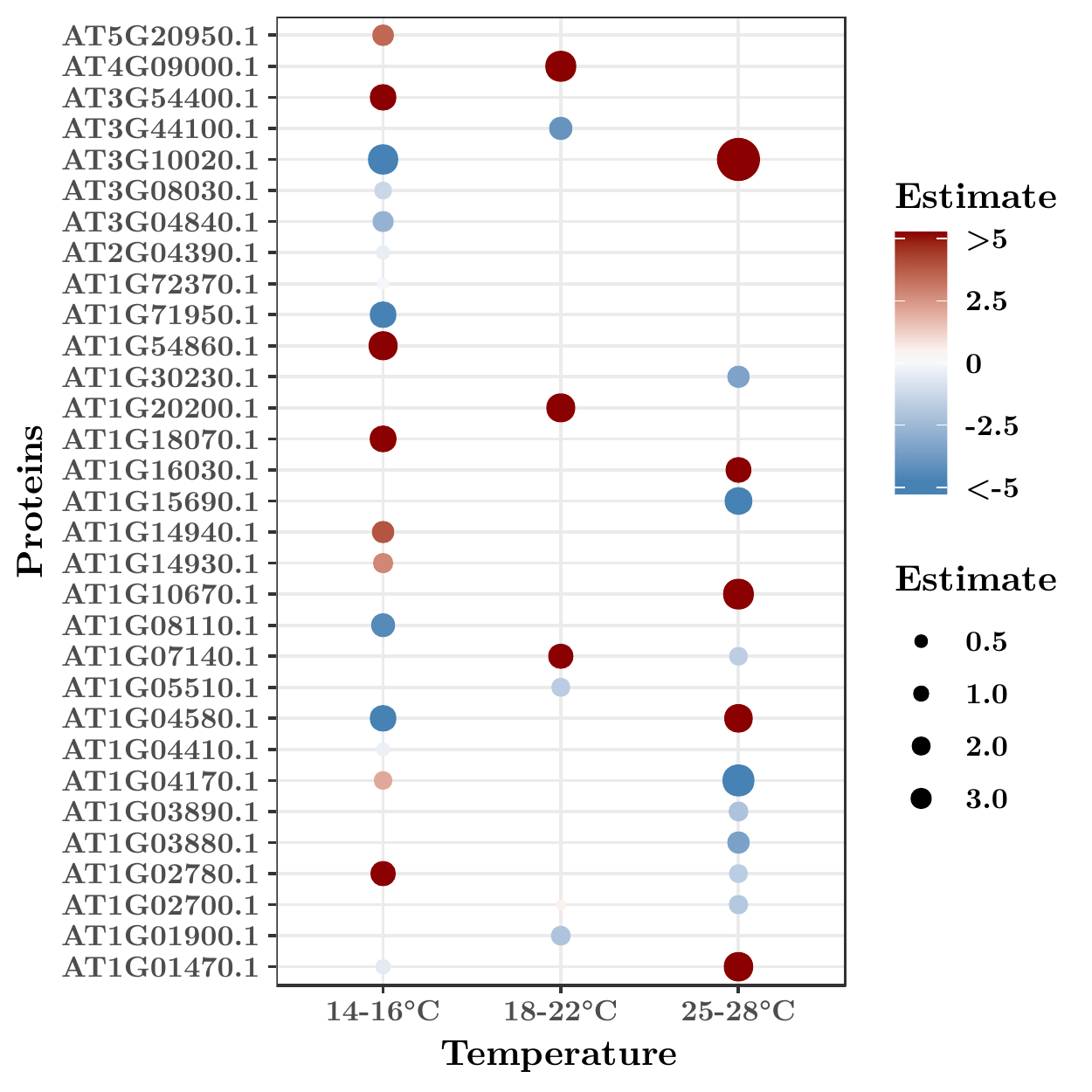}
\caption{Values of the coefficients obtained using the package \texttt{MultiVarSel} with a threshold of 0.95 on the proteomic dataset. \label{fig:proteome}}
\end{figure}

The variable selection in the multivariate linear model using the R package \texttt{MultiVarSel} provided 31 proteins differentially 
accumulated in seeds produced under low, medium or elevated temperature. 
An aspartyl protease (AT3G54400), belongs to both, the Group 8 and to the proteins selected by \texttt{MultiVarSel}. This cell wall associated protein was up-acccumulated in dry seeds 
produced under low temperature. The gene encoding for this protease was described as a cold responsive gene assigned to the C-repeat binding factor (CBF) regulatory pathway, see 
\cite{vogel_2006}. This pathway is requested for regulation of dormancy induced by low temperatures, see \cite{kendall_2011}. Consistently, in Figure \ref{fig:proteome},
two other proteins related to cell wall organization, 
a beta-glucosidase (BGLC1, AT5G20950) and a translation elongation factor (eEF-1B$\beta$1, AT1G30230) were differentially accumulated in seeds produced under contrasted temperature. 
eEF-1B$\beta$1 is associated to plant development and is involved in cell wall formation, see \cite{hossain_2012}. These results suggest that cell wall rearrangements occur under temperature effect during seed maturation. 

As displayed in Figure \ref{fig:proteome}, 6 other proteins involved in mRNA translation: AT1G02780, AT1G04170,
AT1G18070, AT1G72370, AT2G04390 and AT3G04840 were selected.
 The absolute failure of seed germination in the presence of protein synthesis inhibitors underlines the essential role of translation for achieving this developmental process, see \cite{rajjou_2004}. Previous studies highlighted the importance of selective and sequential mRNA translation during seed germination and seed dormancy, see \cite{galland_2014}, \cite{bai_2017} and \cite{bai_2018}. 
Thus, exploring translational regulation during seed maturation and germination through the dynamic of mRNA recruitment on polysomes or either neosynthesized proteome are emerging 
fields in seed research.

\section{Conclusion}

In this paper, we propose a fully data-driven methodology for estimating large block structured sparse covariance matrices in the case where
the number of variables is much larger than the number of samples without limiting ourselves to block diagonal matrices.  
Our methodology can also deal with matrices for which the block structure only appears if the columns and rows are permuted according to an unknown permutation.
Our technique is implemented in the R package \texttt{BlockCov} which is available from the Comprehensive R Archive Network and from GitHub.
In the course of this study, we have shown that  \texttt{BlockCov} is a very efficient approach both from the statistical and numerical point of view.
Moreover, its very low computational load makes its use possible even for very large covariance matrices having several thousands of rows and columns.

\section*{Acknowledgments}

We thank the members of the EcoSeed European project (FP7 Environment, Grant/Award Number: 311840 EcoSeed, Coord. I. Kranner). 
IJPB was supported by the Saclay Plant Sciences LABEX (ANR-10-LABX-0040-SPS). We also thank the people who produced the biological material and the proteomic and metabolomic analysis. In particular, we would like to thank the Warwick University (UWAR, Finch-Savage WE and Awan S) for the production of seeds, the Plant Observatory-Biochemistry platform (IJPB, Versailles; Bailly M, Cueff G) 
for having prepared the samples for the proteomics and metabolomics, the PAPPSO Proteomic Plateform (GQE-Moulon; Balliau T, Zivy M) for mass spectrometry-based proteome analysis and the Plant Observatory-Chemistry/Metabolism platform (IJPB, Versailles; Clement G) for the analysis of GC/MS-based metabolome analyses.

\newpage

\section{Appendix}

\subsection{Variable selection performance}\label{FADA} \hfill

\begin{figure}[!htbp]
\begin{center}
\includegraphics[width=0.9\textwidth]{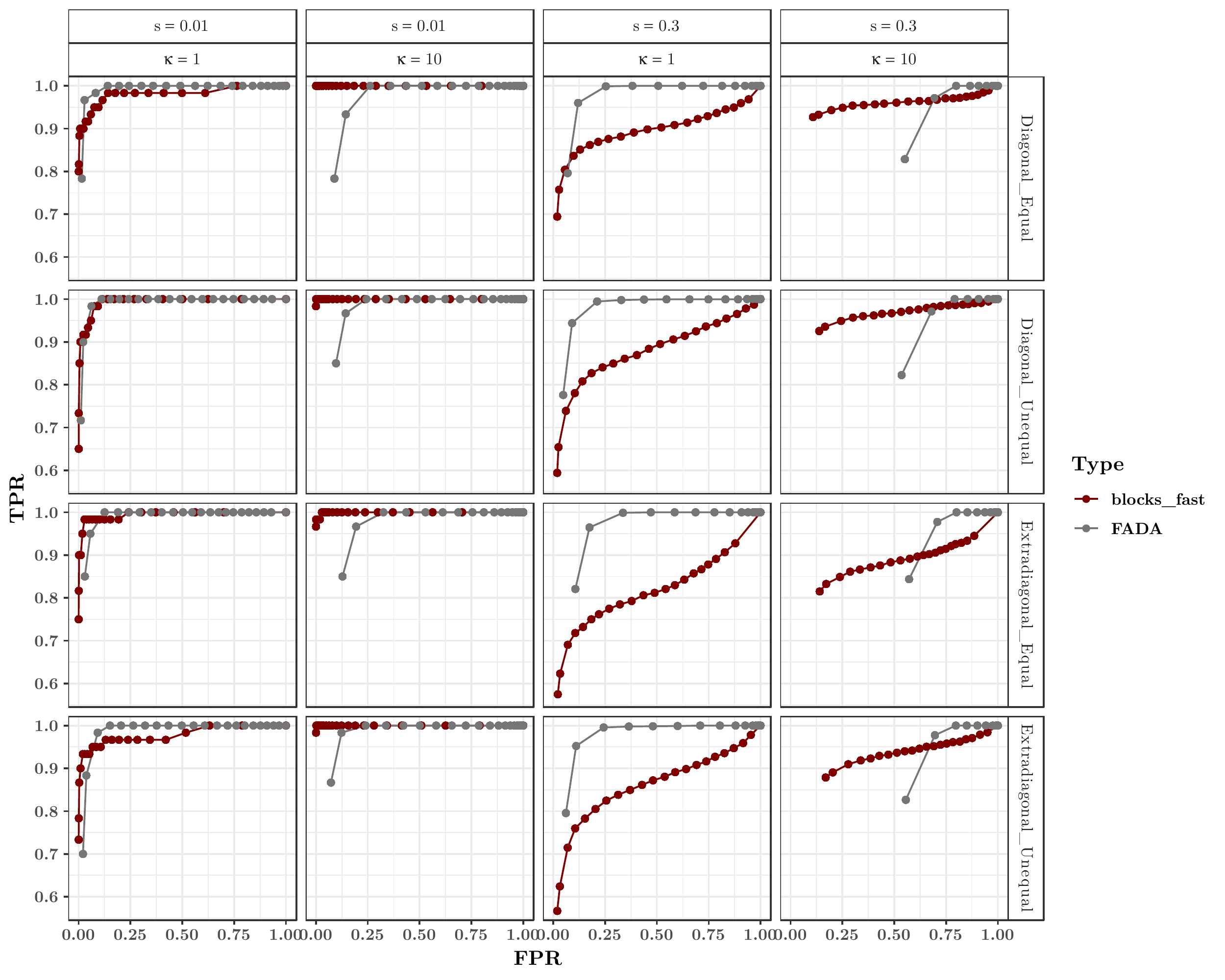}
\caption{Means of the ROC curves  obtained from 100 replications comparing the variables selected by the MultiVarSel strategy using either $\boldsymbol{\Sigma}^{-1/2}$ obtained by BlockCov to remove the dependence or  the methodology proposed by FADA methodology. $\kappa$ is linked to the signal to noise ratio and $s$ denotes the sparsity levels \textit{i.e} the fraction of non-zero elements in $\bB$.\label{fig:comp_fada1}}
\end{center}
\end{figure}

\begin{figure}[!htbp]
\begin{center}
\includegraphics[width=0.9\textwidth]{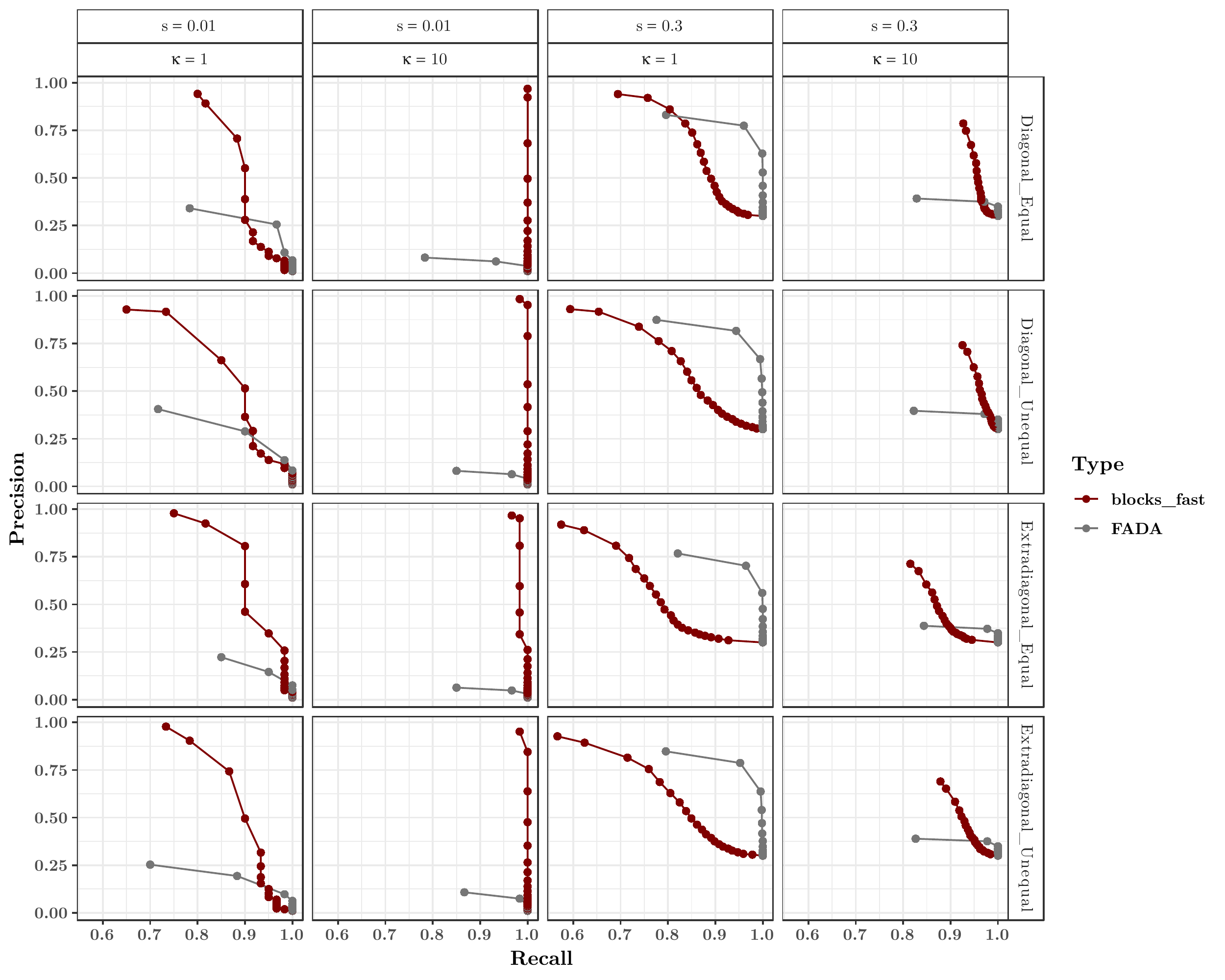}
\caption{Means of the precision recall curves obtained from 100 replications comparing the variables selected by the MultiVarSel strategy using either $\boldsymbol{\Sigma}^{-1/2}$ obtained by BlockCov to remove the dependence or  the methodology proposed by FADA methodology. $\kappa$ is linked to the signal to noise ratio and $s$ denotes the sparsity levels \textit{i.e} the fraction of non-zero elements in $\bB$.\label{fig:comp_fada2}}
\end{center}
\end{figure}
\hfill \\

\newpage 

\subsection{Groups of metabolites}\label{sec:ap} \hfill

\begin{table}[!htbp]
\centering
{\tiny{
\begin{tabular}{llll}
  \hline
Group 1 & Group 2 & Group 3 & Group 4 \\ 
  \hline
Alanine & Arginine & Glutamate & beta.Sitosterol \\ 
  Asparagine & Cystein & alpha.Tocopherol & Campesterol \\ 
  Aspartate & Gaba & gamma.Tocopherol & Eicosanoate \\ 
  Glycine & Glutamine & Linolenic.acid & Heptadecanoate \\ 
  Isoleucine & Tryptophan & H2SO4 & Stearic.acid \\ 
  Leucine & Linoleic.acid & X2.Oxoglutarate & Tetracosanoate \\ 
  Lysine & Quercetin & Mannitol & BenzoylX.Glucosinolate.3 \\ 
  Phenylalanine & BenzoylGlucosinolate.3Breakdown & Urea & Sulfite \\ 
  Proline & Nonanenitrile.9.methylthio & Fructose.6.P & U2609.4.361 \\ 
  Serine & UGlucosinolatebreakdown140.5 & Digalactosylglycerol & U3122.4.202.I3M. \\ 
  Threonine & X2.Hydroxyglutarate & Galactinol & dihydroxybenzoate \\ 
  Tyrosine & Citrate & Galactosylglycerol & beta.indole.3.acetonitrile \\ 
  Valine & Erythronate & Rhamnose & U1837.6.368 \\ 
  X5..methylthio.pentanenitrile & Galactonate & Stachyose & U1841.9.394 \\ 
  Octanenitrile.8.methylthio & Gluconate & Sucrose & U2003.8.293 \\ 
  UGlucosinolatebreakdown140.4 & Glycerate & U1093.6.147 & U2371.1.361 \\ 
  Succinate & Malate & U1124.3.140 & U2375.6.191 \\ 
  Threonate & Allantoin & U1530.2.314 & U2513.2.296 \\ 
  Arabitol & Erythritol & U2053.6.321.1 & U2513.2.296.1 \\ 
  myo.Inositol & Ethanolamine & U2109.3.305 & U2692.9.361 \\ 
  Glycerol.3.P & Sorbitol & U2197.2.494 & U2798.377 \\ 
  myo.Inositol.1.P & Threitol & U2315.2.245 & U2942.2.556 \\ 
  Phosphate & Xylitol & U3898.1.204 & U3063.0.361 \\ 
  U2206.2.299 & Ethylphosphate &  & U3415.9.498 \\ 
  Fructose & Glucose &  &  \\ 
  Glucopyranose..H2O. & Mannose &  &  \\ 
  U1154.3.156 & Raffinose &  &  \\ 
  U1393.172 & Ribose &  &  \\ 
  U1541.8.263 & U1127.5.140 &  &  \\ 
  U1647.2.403 & U1172.9.281 &  &  \\ 
  U1705.2.319.pentitol. & U1559.4.217 &  &  \\ 
  U1729.0.273 & U1628.9.233 &  &  \\ 
  U1816.2.228 & U1849.2.285 &  &  \\ 
  U1859.2.246 & U1927.0.204 &  &  \\ 
  U2076.9.204 & U1939.1.210 &  &  \\ 
  U2170.6.361 & U1983.0.217 &  &  \\ 
  U2184.1.299 & U1983.0.217.1 &  &  \\ 
  U2251.5.361 & U2012.7.361 &  &  \\ 
  U2278.6.361 & U2282.4.349 &  &  \\ 
  U2550.7.149 & U2400.1.179 &  &  \\ 
  U2731.2.160 & U2779.9.361 &  &  \\ 
  U2857.8.342 &  &  &  \\ 
  U2929.1.297 &  &  &  \\ 
  U3041.1.361 &  &  &  \\ 
  U3080.7.361 &  &  &  \\ 
  U3100.8.361 &  &  &  \\ 
   \hline
\end{tabular}

\centering
\begin{tabular}{llll}
  \hline
Group 5 & Group 6 & Group 7 & Group 8 \\ 
  \hline
Quercitrin & X4MTB & BenzoylGlucosinolate.2Breakdown & Maleate \\ 
  Dehydroascorbate & X5MTP & Hexanenitrile.6methylthio & Pentonate.4 \\ 
  Fumarate & X6MTH & Sinapinate.trans & U1408.4.298 \\ 
  Sinapinate.cis & X7MTH & Anhydroglucose & U1617.8.146 \\ 
  Arabinose & X8MTO & U1125.1.140 & U1767.3.243 \\ 
  Galactose & UGlucosinolate140.1 & U1290.198 & U1904.9.204 \\ 
  U1127.4.169 & U1129.9.184 & U1371.5.151 & U2828.8.361 \\ 
  U1718.0.157 & U1270.1.240 & U1549.7.130 & U2839.3.312 \\ 
  U1931.5.202 & U1897.2.327 & U1568.5.313 & U2882.5.297 \\ 
  U2261.0.218 & U2473.361 & U1592.8.217 & U3008.3.457 \\ 
  U2412.1.157 & U2529.8.361 & U1700.6.288 & U3168.2.290 \\ 
  U2588.9.535 & U2756.4.271 & U1759.4.331 & U3218.5.297 \\ 
  U2688.5.333 & U2924.3.361 & U1852.0.217 & U3910.6.597.Trigalactosylglycerol. \\ 
  U3213.1.400 & U3279.7.361 & U1872.1.204.methyl.hexopyranoside. & U2443.7.217 \\ 
  U1380.5.184 &  & U1958.217 &  \\ 
   &  & U2053.6.321 &  \\ 
   &  & U2087.6.321 &  \\ 
   &  & U2150.9.279 &  \\ 
   &  & U2271.6.249 &  \\ 
   &  & U3188.1.361 &  \\ 
   &  & U3701.368 &  \\ 
   &  & U4132.5.575 &  \\ 
   \hline
\end{tabular}
}}
\end{table}
 \hfill
 
\newpage  
 
\subsection{Groups of proteins}\label{sec:protgroup} \hfill

\begin{table}[ht]
\centering
{\tiny{
\begin{tabular}{lr}
  \hline
Proteins & Group \\ 
  \hline
AT1G14170.1 &   8 \\ 
  AT1G20260.1 &   8 \\ 
  AT1G42970.1 &   8 \\ 
  AT1G47980.1 &   8 \\ 
  AT1G55210.1 &   8 \\ 
  AT1G75280.1 &   8 \\ 
  AT2G19900.1 &   8 \\ 
  AT2G22240.1 &   8 \\ 
  AT2G28900.1 &   8 \\ 
  AT2G32920.1 &   8 \\ 
  AT2G37970.1 &   8 \\ 
  AT3G12580.1 &   8 \\ 
  AT3G13930.1 &   8 \\ 
  AT3G26650.1 &   8 \\ 
  AT3G26720.1 &   8 \\ 
  AT3G44300.1 &   8 \\ 
  AT3G47930.1 &   8 \\ 
  AT3G54400.1 &   8 \\ 
  AT3G55800.1 &   8 \\ 
  AT4G16760.1 &   8 \\ 
  AT4G20830.1 &   8 \\ 
  AT4G25740.1 &   8 \\ 
  AT4G34870.1 &   8 \\ 
  AT4G35790.1 &   8 \\ 
  AT5G11880.1 &   8 \\ 
  AT5G12040.1 &   8 \\ 
  AT5G14030.1 &   8 \\ 
  AT5G17380.1 &   8 \\ 
  AT5G22810.1 &   8 \\ 
  AT5G26000.1 &   8 \\ 
  AT5G50370.1 &   8 \\ 
  AT5G66190.1 &   8 \\ 
  AT5G67360.1 &   8 \\ 
  ATCG00480.1 &   8 \\ 
   \hline
\end{tabular}
}}
\end{table}

\newpage

\bibliographystyle{apalike}
\bibliography{biblio_blc}
\end{document}